# Multidimensional Assessment of Public Space Quality: A Comprehensive Framework Across Urban Space Typologies

**Mary John [1], Sherzod Turaev [1], Saja Al-Dabet [1] and Rawad Abdulghafor [2]**

[1]Department of Computer Science and Software Engineering, College of Information Technology, United Arab Emirates University, Al Ain 15551, United Arab Emirates.

[2]Faculty of Computer Studies (FCS), Arab Open University, Oman, Muscat 130, Oman; rawad.a@aou.edu.om

*Correspondence: sherzod@uaeu.ac.ae (S.T)

Abstract: This study presents a comprehensive framework for evaluating the quality of public spaces across various urban typologies. Through a systematic review of 159 research studies, we identify universal quality factors that transcend spatial types as well as specialized factors unique to specific public environments. Our findings establish accessibility (73.6%), safety/security (58.4%), and comfort (52.8%) as foundational requirements across all public space types, while revealing distinct quality priorities for different typologies: open spaces emphasize comfort (70%), parks prioritize activities (60%), green spaces focus on aesthetics and natural elements (70% and 60%), and public facilities uniquely emphasize indoor environment quality (41.7%). The research reveals a hierarchical relationship between factors, where accessibility enables other qualities, safety serves as a prerequisite for utilization, and comfort determines engagement quality. We identify critical limitations in current assessment approaches, including artificial intelligence studies focused on easily quantifiable factors, domain-specific research confined within disciplinary boundaries, and overreliance on subjective perceptions without objective measures. This research provides a foundation for integrated approaches to public space assessment that acknowledge the complexity of public urban environments while addressing both universal human needs and context-specific requirements. The findings support urban planners, designers, and policymakers in developing balanced assessment methodologies that ensure both comparability across spaces and sensitivity to local conditions, ultimately contributing to the creation of high-quality public spaces that enhance urban life and community wellbeing.

## 1. Introduction

A public place is an area where people gather, socialise, and go about their everyday lives, influencing the cultural and social structure of a community. Public spaces include all areas that are publicly owned or available for public use, accessible to everyone for free and without a profit motive [1]. These areas, such as streets, parks, plazas, and other open spaces, function as the city's "open-air living room". Every public space has distinctive spatial, historical, environmental, social, and economic characteristics [1]. They serve as versatile urban assets, linking crucial areas while providing protection, relaxation, and spaces for meetings. Public spaces act as comprehensive integrators, fostering social inclusion, public health, and local economic development [2]. Although facing management challenges and critiques [3], quality public spaces are crucial for attracting investments, residents, and visitors, making them essential for sustainable urban growth and competitiveness [4]. Public spaces should ideally function as inclusive platforms for social interaction, political action, and cultural exchange, reflecting the diverse needs and aspirations of urban communities [2].

Public spaces exist at different levels of urbanization, each having unique characteristics [5]. Urban public spaces in densely populated areas are diverse and abundant, including city parks, plazas, and civic squares serving large populations [6]. Semi-urban spaces in suburban areas [7] or small towns blend urban and rural elements, typically focused around town centers and community parks. Rural public spaces [6], [8] , like village greens and marketplaces, serve as community focal points in less populated areas, often reflecting local traditions and agricultural heritage. Despite these differences, public spaces at all urbanization levels share a core purpose: fostering social interaction, community gatherings, and civic engagement [9]. These spaces adapt to local needs and resources in bustling cities or quiet villages [10]. Their design and function reflect the specific context of their communities, highlighting the versatility of public spaces in supporting social cohesion across diverse settings [11], [12]. This adaptability underscores the universal importance of public spaces in enhancing community life, regardless of urbanization level [13], [14].

*1.1. Background*

Examining specific spatial types within public spaces provides a detailed view of their forms and functions. Public spaces can be grouped into three main categories: streets, open spaces, and public facilities [2]. Both walkable and drivable streets form the connective tissue of public space. They include pedestrian-friendly areas, squares, traffic islands, and tramways, facilitating movement and social interaction [15]. Open spaces, characterized by the absence of buildings, include parks, green areas, and waterfronts [16]. These provide relief from urban density and create opportunities for recreation and connection with nature. Public facilities like libraries, civic centers, markets, and sports venues act as built environments that anchor community life and promote civic engagement [1], [2]. Each spatial type adapts to the specific needs of urban, semi-urban, and rural contexts while maintaining its core function of supporting social interaction and community vitality. These spatial types collectively shape the public realm, with each making a unique contribution to the social fabric of communities. Their diverse functions highlight the significance of varied public spaces in creating vibrant, livable environments across different levels of urbanization.

The various types of public spaces play a crucial role in shaping community life and enhancing individual well-being. Their significance goes beyond physical features, affecting diverse aspects of human experience. Public spaces contribute to overall well-being by providing opportunities for physical activity, social interaction, and connection with nature. Parks and green spaces offer areas for exercise and relaxation, helping to reduce stress and improve both mental and physical health [17]. Streets and squares foster social connectivity by hosting planned events and spontaneous encounters. This social aspect is essential for mental well-being, especially in combating isolation in urban environments [18].

Public facilities like libraries, civic centers, and sports venues stimulate cognitive development through exposure to diverse experiences and cultural activities. They also serve as forums for public discourse and civic engagement, supporting democratic participation [1], [2]. These spatial types, function as community "living rooms", building social capital and enhancing livability across urban, semi-urban, and rural settings. Collectively, these spaces form the backbone of community life, promoting public health, social cohesion, and urban vitality [19], [20], [21]. Their multi-faceted impact underscores the need for thoughtful design and maintenance of public spaces in all community contexts.

The quality of public spaces significantly influences urban life and the well-being of communities. While grasping the various types and definitions of public spaces offers a foundational perspective, examining the factors that enhance their effectiveness, and value is just as crucial. These factors encompass a broad range of considerations. Comfort, for instance, encompasses both thermal and environmental aspects. It is about creating spaces where people want to linger, protected from harsh weather and surrounded by pleasant sensory experiences. Accessibility is another critical factor – a truly public space should be easy to reach and welcoming to all, regardless of physical ability or socioeconomic status. Other important factors include publicness (the degree to which the space feels open and available to all), lighting (for safety and ambiance), mobility (ease of movement within the space), inclusiveness (how well it serves diverse populations), emotional perceptions (the feelings the space evokes), and walkability (the ease and pleasure of pedestrian movement). These factors do not exist in isolation but interact in complex ways to shape the overall quality of a public space. These factors can be viewed through various lenses or dimensions: social, cultural, physical, environmental, economic, psychological, functional, and managerial. Each dimension offers a unique perspective on how public spaces operate and what they mean to communities. Identifying and understanding these factors is crucial for effective evaluation and improvement of public spaces. High-quality public spaces are more than just nice-to-have amenities; they are essential components of healthy, vibrant communities. They provide places for people to interact with nature and each other, directly impacting mental health and well-being. Moreover, the quality of public spaces can be seen as a reflection of a society's values and priorities. For the younger generation, high-quality public spaces offer opportunities for physical activity and social interaction, crucial for healthy development. These spaces can also serve as incubators for ideas and creativity, where people come together to exchange thoughts and develop new concepts. The impact of public spaces on neighborhood health and well-being cannot be overstated. A truly 'public' space should be easily accessible, comfortable to use, and facilitate easy movement. This means considering factors like temperature, lighting, seating, and amenities, as well as ensuring proper sidewalks and walkable areas.

*1.2. Need for review*

Despite extensive research into individual aspects of public-space quality, no single review brings together all typologies under a unified lens. Most syntheses limit themselves to one or two dimensions social life ([22]), inclusivity ([23]), quality of life ([13]), or walkability and sustainability ([24]) while domain-specific work remains equally narrow: green-space evaluations ([25], [26]) exclude streets and plazas; street-level studies focus on comfort ([25]) or thermal conditions ([26]); park research centers solely on perceived restorativeness([27]). The resulting methodological patchwork from GIS analyses to psychometric surveys prevents meaningful comparison or integrated framework development.

This narrative review addresses that gap by examining domain-based studies across streets, plazas/waterfronts, parks/green areas, and civic facilities, identifying quality factors within seven dimensions (social, cultural, physical, environmental, functional, economic, managerial), and proposing a visual framework for more holistic, context-sensitive public-space evaluation.

*1.3. Objective*

To frame the direction and deliverables of this narrative review, we set forth the following objectives:

1. Synthesize 159 domain-based studies of publicly accessible urban spaces (streets, plazas/waterfronts, parks/green areas, civic facilities) into a cohesive narrative.
2. Identify and catalog every quality factor these studies employ, organizing them under seven dimensions: social, cultural, physical, environmental, functional, economic, and managerial.
3. Quantify the frequency of each factor's use across the four space typologies.
4. Map factor–typology relationships and present their hierarchies using visual aids.

*1.4. Review Scope*

This narrative review focuses exclusively on domain-based studies of permanently accessible, non-commercial urban spaces namely streets (pedestrian and vehicular), plazas and waterfronts, parks and green areas, and civic facilities (libraries, markets, sports venues). We deliberately exclude AI-driven or machine-learning papers from our main corpus; these are only mentioned in our discussion to highlight that such approaches typically consider a limited subset of quality factors and often omit dimensions critical to space types.

Drawing on 159 peer-reviewed articles from urban planning, environmental psychology, public health, architecture, and related fields, we employ a seven-dimensional framework: social, cultural, physical, environmental, functional, economic, and managerial, to map the constituent factors (e.g., accessibility, safety, comfort, inclusivity, activities, natural elements, maintenance). For each typology, we chart factor prevalence and hierarchical relationships (e.g., accessibility as an enabler, safety as a prerequisite, comfort as a driver of sustained use).

By narrowing our lens to narrative, mixed-methods, and qualitative evaluations in domain-based literature, we ensure a holistic, user-centered synthesis. In the discussion, we note that emerging AI-based assessments, while promising for scalability tend to rely on easily quantifiable metrics and therefore miss context-specific factors essential for comprehensive public-space quality appraisal.

## 2. Methodology

This section describes the four-step approach used to identify, select, analyze, and assess domain-based studies of public-space quality.

*2.1. Search Strategy*

A comprehensive search was carried out in seven academic databases: Scopus, Web of Science, Google Scholar, PubMed, IEEE Xplore, Semantic Scholar, and SciSpace using a combination of broad and targeted keywords, including "public space quality", "urban public spaces", "public space assessment", "comfort", "health and safety", "publicness", "walkability", "environmental factors", "accessibility", and "inclusiveness". No restrictions were imposed on publication year, thereby capturing both foundational and recent work, while the search was limited to English-language records appearing in peer-reviewed journals, conference proceedings, books, or technical reports.

Although this paper presents a narrative review rather than a systematic review, we employed a structured search and screening process to ensure comprehensive coverage of relevant literature, as illustrated in Figure 1. From an initial 489 records identified through database searches, 277 were excluded during title and abstract screening. These excluded records primarily consisted of policy documents without empirical evaluation (n=85), prescriptive design guidelines lacking quality assessment (n=67), and papers with contexts irrelevant to public space quality (n=125). The remaining 212 full-text articles were assessed for eligibility against our inclusion criteria.

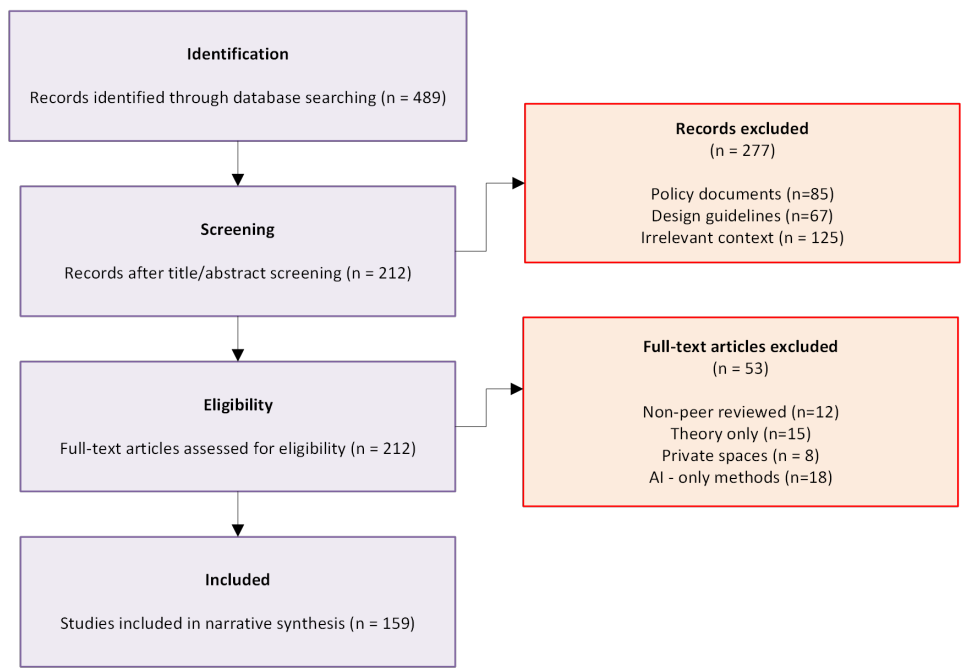

**Figure 1.** Study selection process.

During full-text review, we excluded an additional 53 papers: non-peer reviewed materials such as reports or white papers (n=12), studies presenting purely theoretical discussions without empirical assessment (n=15), research focused on private or semi-public spaces rather than true public environments (n=8), and studies using exclusively AI or machine learning methods without user-centered evaluation components (n=18). This structured screening process resulted in a final corpus of 159 domain-based studies that formed the foundation for our narrative synthesis.

The quality assessment applied during screening emphasized clear research objectives, appropriate sample sizes, valid assessment tools, and rigorous empirical data collection methods. This approach ensured that our analysis was based on methodologically sound studies while still allowing for the interpretive flexibility characteristic of narrative reviews. By combining the strengths of structured screening with narrative synthesis, we provide reliable insights into public space quality across diverse urban contexts while facilitating a coherent narrative that highlights patterns, connections, and thematic developments in the literature.

*2.2. Inclusion and Exclusion Criteria*

The full texts of the 212 remaining papers were then assessed against predefined inclusion and exclusion criteria. Studies qualified for inclusion if they presented empirical or mixed-methods evaluations of one or more public-space quality factors such as comfort, safety, inclusiveness, or publicness and focused on streets, plazas/waterfronts, parks, green areas, or civic facilities. Studies were excluded if they were non–peer-reviewed, confined to theoretical or policy discussions without explicit quality measurement, centered on private or semi-public environments (e.g., gated communities or commercial malls), or relied solely on AI or machine-learning techniques without a user-centered evaluation focus. Application of these criteria yielded a final corpus of 159 domain-based studies.

## 2.3. Analysis Framework

Each paper in the final corpus was first subjected to a narrative integration, wherein its context, objectives, and methodological approach were reviewed in full to understand how public-space quality had been defined and evaluated. Building on this synthesis, all explicitly reported quality factors were extracted carefully noting the terminology used, the typology of space studied, and the methods of measurement employed. These factors were then organized into seven analytical dimensions (social, cultural, physical, environmental, functional, economic, and managerial) and mapped across five typologies: streets; plazas and waterfronts; parks; green areas; and civic facilities to determine their relative prevalence.

Patterns emerging from this mapping were visualized through radar charts, which revealed each typology's distinctive "fingerprint" of quality dimensions. These visual representations informed a richer interpretive narrative, illustrating how disciplinary traditions and methodological choices shape factor selection. This approach enabled us to identify both common quality factors across all space types and distinctive priorities unique to specific public space typologies.

## 2.4. Quality Assessment Criteria

A stringent screening protocol underpinned the selection of studies for analysis, ensuring that only those with robust empirical evaluations were retained. Each paper was required to describe measurement methods clearly whether through surveys, observational protocols, or environmental metrics and to employ defined frameworks or measurable indicators directly linked to outcomes such as usability, safety, comfort, or inclusiveness. Studies lacking explicit assessment criteria or confined to design guidelines or policy discussion were excluded. This rigorous filtering produced a coherent, high-quality corpus that supports both narrative and quantitative synthesis of public-space quality factors.

To minimize selection bias, a structured quality assessment protocol focusing on methodological rigor and relevance was employed. Studies were evaluated based on: (1) Clarity of research objectives and methods; (2) Appropriateness of sample size and selection; (3) Validity and reliability of assessment tools; (4) Robustness of data analysis; and (5) Relevance to public space quality assessment. Geographic representation was also considered, with efforts made to include studies from diverse urban contexts across six continents, though we acknowledge a predominance of research from European (36%), Asian (31%), and North American (22%) contexts, with fewer studies from South America, Africa, and Oceania (11% combined). This distribution reflects current publication patterns in the field but may limit generalizability to some regional contexts. Studies with purely theoretical frameworks, insufficient methodological details, or very small sample sizes (n<20 for user perception studies) were excluded to maintain empirical quality.

# 3. Conceptual Frameworks and Theoretical Models

Public spaces are essential components of urban environments, contributing significantly to the quality of life, social cohesion, and sustainability of cities. Over the years, researchers and urban planners have developed various conceptual frameworks and theoretical models to assess and improve the quality of public spaces. This section explores these frameworks, highlighting their key components, methodologies, and applications in diverse urban contexts. At the core of these frameworks lie several fundamental aspects of public space quality, which have evolved and expanded over time to encompass a more comprehensive understanding of urban dynamics. The fundamental aspects of public space quality often revolve around accessibility, safety, comfort, and functionality. However, as the understanding of urban dynamics continues to develop, the frameworks for evaluating public spaces are also adapting to incorporate these evolving perspectives. [1] emphasizes eight quality factors: accessibility, safety and security, comfort and image, uses and activities, sociability, environmental sustainability, economic value, and cultural and historical significance. This comprehensive approach sets the stage for more sophisticated evaluations of public spaces.

## 3.1. Evolution Of Public Space Quality Assessment

Building upon these foundational concepts, researchers have developed precise, quantifiable methods and tools for assessing public space quality based on accessibility, safety, comfort, and functionality. For instance, [23] introduces the Public Space Index (PSI), which measures public spaces across five key dimensions: inclusiveness, meaningful activities, safety, comfort, and pleasurability. When applied in Tampa, Florida, this framework revealed that while spaces were comfortable, safe, and inclusive, they often lacked meaningful activities and pleasurability. Similarly, [28] presents a validated tool with 39 items across six dimensions (comfort and relaxation, safety and security, accessibility, belonging to space, social interactions, and discovery and learning), demonstrating high reliability and validity when applied in Sana'a, Yemen. Although these comprehensive frameworks lay a strong foundation for public space evaluation, the diversity of urban environments necessitates more adaptable approaches. [7] addresses this need by proposing a framework based on five key design principles: accessibility, specificity, authenticity, adaptability, and functionality. This approach emphasizes the importance of context-sensitive design and quality evaluation in creating successful public spaces, particularly for small parks that play a crucial role in providing accessible nature experiences in urban areas.

## 3.2. Key Theoretical Frameworks

Established frameworks continue to offer valuable guidance, but recent technological advancements are reshaping public space assessment, marking the integration of technology as an emerging trend. [29] proposes a "smart public space" concept with six layers: governance, physical infrastructure, data collection, communication, server infrastructure, and smart services. This approach aims to enhance inclusivity through technology and citizen participation. Similarly, [30] emphasizes the integration of advanced technologies like remote sensing and GIS in assessment methodologies. While technological advancements offer overarching solutions, specialized frameworks have been developed to address the unique challenges of specific public space typologies, with some focusing on types of public spaces. [31] develops a conceptual framework for assessing underground space quality, integrating functional, psychological, and structural aspects. [32] presents a framework for assessing urban green spaces based on availability, accessibility, and characteristics. It proposes a standardized GIS-based indicator suggesting that urban residents should live within 300m of green space. This aligns with the recommendations in [33], which provides a comprehensive framework for planning, designing, and managing urban green spaces, emphasizing factors such as environmental quality, health, social cohesion, and community engagement. Beyond the physical and technological aspects of public spaces, researchers have also begun to explore the ethical considerations involved in their design and allocation, as addressed in [34], which presents a framework of allocation mechanisms derived from ethical principles and normative perspectives on street use. This approach provides a structured way to consider ethical implications in street design and space allocation.

While ethical considerations provide a crucial foundation for equitable public space design, the ultimate success of these spaces hinges on how well they meet the needs and preferences of their users. This recognition has led to the development of frameworks that prioritize human-scale considerations and user perspectives, underscoring their importance in creating effective and inclusive public spaces. [35] presents a novel approach to measuring human-scale living convenience, using buildings as analytical units. [36] proposes a conceptual framework linking use patterns, needs, preferences, and quality of neighborhood parks, emphasizing the importance of considering users' perspectives in designing and managing urban green spaces. Though human-scale and user-centric approaches offer valuable insights into individual spaces, urban planners and policymakers increasingly recognize the necessity for more holistic, city-wide strategies to ensure the equitable distribution and connectivity of public spaces throughout urban areas. Several frameworks address this need for comprehensive, city-wide approaches, highlighting the importance of a coordinated effort in urban planning. [4] and [37] offer systematic approaches to assessing public spaces across entire cities, focusing on network, distribution, accessibility, quantity, and quality. These tools emphasize social inclusion and stakeholder participation throughout the assessment process [2] presents a tripartite approach to public space strategy development, emphasizing the importance of viewing public spaces as an interconnected system. Even though city-wide approaches offer valuable strategies, the diverse nature of urban environments worldwide necessitates the development of specialized frameworks, to address the unique needs of urban settings. [38] presents a framework for assessing and improving public spaces in large cities, identifying ten key quality indicators and emphasizing a multi-scale approach. [20] introduces the Public Space Quality Index (PSQI), which is particularly relevant for developing countries where public space conditions may differ significantly from those in developed nations.

Even though many frameworks offer complex, quantitative approaches to public space assessment, simpler, more intuitive methods also provide valuable insights and can be easily applied across various contexts. These qualitative approaches play an important role in public space assessment, complementing the more intricate frameworks.

### 3.3. Gaps In Current Frameworks

Existing public-space quality assessments offer a variety of tools from quantitative indices to qualitative checklists and human-scale guidelines, but they tend to concentrate on a limited set of factors or a single typology. Reviews may catalog comfort, walkability, or inclusiveness in isolation, or focus solely on parks, streets, or plazas, without examining how these factors overlap or differ across space types. As a result, the field lacks a coherent synthesis that brings together all empirically tested quality factors and highlights their commonalities.

The present paper addresses this gap by conducting a narrative review of domain-based studies to identify and group every quality factor used in empirical assessments of streets, plazas/waterfronts, parks, green areas, and civic facilities. Rather than proposing a full unified framework at this stage, the focus here is on cataloging and organizing those factors into seven thematic dimensions social, cultural, physical, environmental, functional, economic, and managerial as a foundation for deeper analysis. By mapping out which factors appear most frequently and where, this work lays the groundwork for a subsequent paper that will develop and validate a comprehensive, cross-typology framework for public-space quality evaluation.

## 4. Public Space Typology and Quality Factors

This section categorizes public spaces based on their spatial and functional characteristics and examines the factors that influence their quality. Public spaces are diverse in form and purpose, ranging from urban streets and open spaces to green areas, parks, and public facilities. Each type serves different roles, and the factors affecting their quality vary accordingly. The following subsections classify public spaces into different typologies and analyze the factors considered in their assessment.

### 4.1. Urban Spaces

As observed, the field of public space quality assessment encompasses a wide range of approaches, from highly urban spaces are places where people interact, exchange ideas, relax, and spend significant amounts of time. To maintain quality, these spaces require continuous monitoring and timely actions to address evolving needs. Quality assessment spans multiple aspects, including physical, environmental, functional, psychological, economic, and managerial perspectives. Studies analyze urban public spaces from various dimensions such as Social, Political, and Environmental [12], while some studies have analyzed and assessed urban spaces, often focusing on specific spatial types or emphasizing certain quality factors.

*Accessibility*: The assessment of accessibility, walkability, and pedestrian spaces has become essential in evaluating the quality of public urban environments. These studies provide insights into how urban design influences pedestrian movement, safety, and overall accessibility within public spaces, creating a foundation for more inclusive and user-friendly urban landscapes. Accessibility in Tehran, Iran is examined in [39], focusing on both objective and subjective perspectives. The research highlights significant discrepancies between physical measures of accessibility (such as distance) and residents' subjective perceptions, indicating that both approaches are vital in shaping accessible public spaces. By balancing objective data with user experiences, this study reveals the complex nature of accessibility and the need for urban planners to integrate both dimensions in design decisions. While [39] centers on accessibility as experienced by users, [40] emphasizes spatial optimization for practical, accessible design. This study introduces a decision support system aimed at improving pedestrian accessibility within neighborhood settings, incorporating public spaces such as green areas, local shops, and recreational facilities within a walkable radius. This structured approach allows planners to make informed decisions on facility placement, optimizing neighborhood layout to enhance walkability. [41] shifts the focus to spatial relationships, using Space Syntax analysis to evaluate pedestrian accessibility. Unlike the decision support system in [40], this approach uses spatial data to reveal how the design and connections within public spaces shape walkability. Key quality factors here include spatial configuration, attraction of opportunities, and pedestrian comfort, illustrating the role of spatial layout in enhancing accessibility and user experience.

Further refining pedestrian space evaluation from accessibility point of view to the pedestrian spaces and routes, [42] employs qualimetric methods to assess specific pedestrian routes. Unlike previous studies that address broader aspects of urban accessibility, this research provides a focused analysis at the path level, identifying route features that can be enhanced to improve usability and pedestrian experience. By examining physical characteristics, safety, aesthetic value, and usability across 50 routes, this study produces a detailed quality classification map that can guide targeted improvements. On a global scale, [43], [44] introduces a standardized index for assessing pedestrian spaces across cities worldwide. The index considers safety, convenience, attractiveness, and policy support, providing a versatile framework adaptable to various local contexts. This global perspective contrasts with the localized, case-specific methods of previous studies, offering a universal tool for evaluating and improving walkability across diverse urban environments.

An important quality factor that is closely related to accessibility and pedestrian spaces is walkability. [45] applies the Walkability Index with an emphasis on elderly health, evaluating pedestrian environments within a historic city center. The study considers urban tissue, urban scene, and safety, finding that only certain paths are truly accessible for older adults. Factors such as steep slopes were found to significantly impact pedestrian access, underscoring the need for targeted adaptations to enhance walkability for elderly users. This age-focused approach adds a valuable dimension to walkability assessments by addressing the specific needs of older populations. Another comprehensive tool for assessing urban walkability is introduced in [46] focusing on factors like modal distribution, urban grid, urban scene, safety, and environmental features. Although the Global Walkability Index provides a standardized assessment, the Walkability City Tool allows for detailed, context-specific analysis, assisting urban planners in making targeted improvements to enhance pedestrian experiences at a micro level. These studies highlight the importance of accessibility, walkability, and pedestrian-friendly design in public spaces. By combining objective metrics, subjective perceptions, spatial analysis, global indices, and age-sensitive adaptations, these approaches contribute to a comprehensive understanding of how to create accessible and inviting urban environments for diverse user groups.

*Mobility:* The relationship between mobility and public space quality has garnered increasing attention, particularly as urban areas prioritize sustainable transport solutions. The following studies highlight how mobility initiatives, like Light Rail Transit (LRT) and pedestrian-friendly designs, impact urban quality, fostering a balance between accessibility and sustainability. It explores the link between sustainable mobility and urban space quality in Granada. This study found that while the introduction of an LRT system improved public transport quality and increased pedestrian areas along its corridor, the surrounding urban fabric still heavily favors motorized vehicles. This outcome illustrates the challenges of shifting towards sustainable mobility within urban areas structured around vehicle traffic. Similarly, [47] assesses the effects of Granada's LRT system on urban mobility and public space quality. However, this study expands the analysis to include seven quality factors: urban activity and vitality, urban image and identity, urban form and design, environmental quality, social cohesion, accessibility and mobility, and safety and security. Improvements in pedestrian areas and overall urban quality along the LRT route highlight the transit system's positive impact on the public realm.

Nevertheless, the findings echo those of [48], noting that much of the city's infrastructure remains tailored to motorized traffic, underscoring the limitations of isolated mobility improvements within car-dominant environments.

A methodology for evaluating public spaces with an eye toward promoting sustainable mobility is introduced in [49]. This study highlights quality factors such as public space quality, safety, and accessibility, identifying specific issues like a lack of cycle paths, unsafe pedestrian crossings, and high vehicular traffic. The findings emphasize the need to integrate technical analyses with user perceptions to develop public spaces that support sustainable mobility, bridging the gap between practical infrastructure and user experience. This analysis is extended in [50] to emphasize a comprehensive approach to urban design, examining sustainable mobility alongside the aesthetic and cultural value of public spaces. This study considers five primary quality factors: functionality and accessibility, safety and security, environmental quality, aesthetic and cultural value, and social interaction and inclusivity. By focusing on these areas, the research underscores the importance of designing public spaces that integrate sustainable mobility with visual appeal and cultural significance, thereby enriching both mobility and quality of life in urban settings. Thus, there exists a complex interplay between mobility and public space quality. Through varied methodologies and focus areas, they demonstrate the need for urban design strategies that prioritize both sustainable mobility and high-quality public spaces to create accessible, safe, and visually appealing environments for diverse urban populations.

*Perception and Emotion:* The perception and emotional impact of public spaces are essential aspects of urban quality, influencing how users experience and connect with their surroundings. Studies in this area highlight the varied factors that affect perception, including built environment elements, user groups, and the emotional responses they elicit. A data-driven analysis of public space perception is presented in [51] using large datasets to quantify subjective responses. It considers six primary quality factors: beautiful, boring, depressing, lively, safe, and wealthy, revealing that specific urban features, particularly vegetation, significantly enhance perceptions across these dimensions. The findings emphasize how the natural environment positively influences public space quality. A similar data-focused approach is seen in [52], which also explores how perceptions vary based on spatial characteristics. Focusing on accessibility, comfort, socialization, and activity as quality factors, this study uncovers that management practices exert a significant influence on perceived quality, often more so than ownership (public vs. private). Perception and emotional responses are thus seen to be crucial to understanding the quality of public space. By combining large datasets, comparative analyses, and advanced technological methods, this body of research emphasizes the need for urban designs that align with user perceptions, fostering positive and emotionally resonant experiences in public spaces.

*Environment and comfort:* People's perception and emotional responses in urban public spaces are closely related to environmental quality and comfort, as these aspects shape how welcoming and engaging the space feels to them. Subjective environmental perceptions are important to measure because they capture the individual and community-level experiences that influence how people feel and behave within public spaces. To measure subjective environmental perceptions in public spaces, a psychometric scale has been developed that assesses two main quality factors, the affective domain and the cognitive domain, as introduced in [53]. Findings reveal that this scale effectively differentiates between types of public spaces and predicts perceptions related to restorativeness, safety, and visitability, underscoring the importance of subjective perceptions in evaluating environmental quality. Understanding how specific environmental characteristics influence users' sense of restorativeness is essential for designing spaces that reduce stress, support mental well-being and increase user satisfaction. [27] explores this relationship by analyzing factors like space design, maintenance (space care), social interaction, and sensory elements. The research finds that parks rate higher in perceived environmental quality and restorativeness than squares, illustrating how design, upkeep, and sensory experiences can enhance the mental restoration potential of public spaces.

Moving beyond individual qualities (safety, comfort, or aesthetic appeal) in user experience, a broader analysis reveals the true significance of public spaces: they are communal assets that foster social interaction, cultural identity, and inclusivity. [54] examines this comprehensive perspective. It examines five core factors: communication and social inclusivity, civic engagement, collective memory, organizational creativity, and recreation. Findings indicate that while public spaces in Irkutsk possess significant physical potential, they often fall short in meeting social inclusion and civic engagement needs, highlighting a gap between structural attributes and social utility. A similar finding is shown in [55], which reveals that conventional indices often prioritize structural aspects, while citizens frequently value comfort, sociability, and design elements, underscoring the need for user-centered assessments. This emphasizes a gap between standardized environmental indices and citizens' preferences. In this paper, the following seven main components are discussed: landscape elements, landmarks and architecture, mobility and infrastructure, health, comfort, and safety, elements of public space and street furniture, activities and sociability, and amenities. Together, these studies illustrate that environmental quality in public spaces involves a complex array of physical and experiential factors.

Thermal environment represents a critical subcomponent of environmental quality. Public spaces undergo various structural and other adjustments to enhance thermal comfort. [56] investigates the contribution of greening and high-albedo coatings to improvements in the thermal environment in complex urban areas. The study examines four main quality factors: thermal environment, green areas, surface reflectivity, and urban geometry. Their findings indicate that green space development is most effective in reducing temperatures in areas with high density of artificial structures. The acoustic environment constitutes another significant contributor to environmental quality. Acoustic environment along with functionality, attractiveness, user perception, and spatial characteristics is analyzed in [57]. Both thermal and acoustic factors determine environmental quality, which directly relates to user comfort. In [58] environmental air quality and noise pollution is analyzed which directly relates to the quality of life. [59] proposes a comprehensive approach to identify qualitative and quantitative criteria that define a comfortable urban space, examining two main quality factors: thermal comfort and psychological adaptation. Their findings emphasize the importance of creating diverse, adaptable, and natural urban environments that enable users to control their thermal experience. Similarly, [25] examines heterogeneity in outdoor comfort assessment in urban public spaces. The study analyses five main quality factors: thermal comfort, environmental perception, psychological factors, behavioral adaptation, and socio-demographic characteristics. Their research identified two distinct latent classes with different causal structures in comfort assessment. Comfort emerges as a user-centered concept that can be maximized through consideration of user preferences. [60] presents a method for controlling devices to maximize comfort levels in public smart spaces, focusing on five main quality factors: environmental comfort, user preference adaptability, context awareness, device control efficiency, and user interface. Their findings demonstrated that the proposed method could accurately estimate user comfort levels while rapidly adapting to changing user preferences.

*Livability:* At the foundation lies livability, where environmental comfort including thermal and acoustic factors, combined with basic amenities to create spaces suitable for human use. One such environmental condition, winter, and its effects on livability are studied in [61]. The study considers seven main quality factors: climate adaptation, accessibility and mobility, social interaction, comfort and shelter, aesthetics and design, functionality, and safety. Findings reveal that, while urban planning incorporates winter conditions, a systematic approach to addressing winter-related challenges is lacking. This highlights the need for climate-adaptive designs to improve year-round livability in public spaces, especially in regions with severe seasonal variations. Another study measuring public space livability is explored in [14], focusing on the role of public benches and their impact on quality of life. Analyzing three quality factors: objective characteristics, subjective characteristics, and aesthetics, the study finds that the design and placement of benches often fall short of user needs, with users prioritizing comfort and convenient proximity over visual design and security. This study emphasizes the importance of integrating basic amenities that align with users' daily needs to create more liveable and user-centered public spaces. From these studies, livability depends on climate adaptability and comfort but also on thoughtful design and the inclusion of essential amenities that enhance daily use. Livability and vitality go hand in hand as quality factors of a public space. [62] evaluates urban quality and vitality in a historical district by considering five quality factors: urban vitality, functional quality and diversity, social cultural quality and city safety, quality and attractiveness in urban space, and local governmental services. The study found positive changes in urban vitality in some neighborhoods, while others experienced negative changes. It emphasizes the importance of functional diversity and traditional urban patterns in maintaining urban vitality. Thus, beyond livability and vitality, inclusiveness is essential for creating public spaces that are accessible and welcoming to everyone, ensuring all users feel valued and supported in the environment.

*Inclusiveness:* Inclusiveness means that people belonging to different groups whether physical needs, age or gender will find the public space useful for their needs. [63] introduces a tool for assessing city inclusiveness for people with disabilities. It considers 20 main quality factors, including public spaces and built environment, transportation and urban mobility, recreation, sports, and leisure. The study found varying levels of inclusivity across different urban domains, with transportation and urban mobility often scoring poorly. Another study in [64], was done from a gender-inclusive perspective. It examines four main quality factors: comfort, image, usage, and vitality. The study found significant differences in preferences between men and women, with women prioritizing comfort (especially safety) and image while men prioritizing vitality and usage. These studies show that inclusiveness in public spaces means meeting diverse needs, from enhancing mobility for people with disabilities to ensuring comfort, safety, and vibrant use for all genders. In addition to inclusiveness, the public focuses on designing spaces that encourage social interactions, active use, and a strong sense of community, fostering a more vibrant and connected urban environment. Publicness in urban spaces includes accessibility and the ability to support diverse social activities and interactions.

*Publicness:* A comprehensive framework for evaluating public space publicness is presented in [65]. It focuses on four main quality factors: activity, physical design, space-user connection, and management. The study emphasizes the need for a transdisciplinary approach to public space analysis and highlights the changing nature of public spaces in contemporary cities. Publicness of different urban space types is evaluated in [66] with a slightly different set of quality factors by defining the Publicness Dimension. They are social life and socialization, activities, access and linkage, and identity and image. The study found that ownership alone does not determine the use or preference of urban spaces, with privately owned spaces like shopping malls sometimes being highly attractive and well-used. This highlights the need for a transdisciplinary approach to measure publicness, providing a comprehensive tool to assess the usage and effectiveness of public spaces.

Accurately quantifying the usage of public spaces provides valuable insights into how these areas serve communities and meet diverse needs. A tool that focuses on measuring Intensity and Diversity of Use (IDU) to assess public space quality, considering factors like physical settings, atmosphere quality, visual appeal, and cultural meanings, is explored in [67]. Findings reveal a significant positive correlation between IDU and overall public space quality, highlighting that spaces with diverse, well-facilitated activities attract greater and more varied usage.

Thus, the importance of advanced tools in understanding usage patterns allows urban planners to optimize public spaces for both intensity and diversity of use, thereby enhancing their relevance and functionality for all users. High levels of usage in public spaces naturally lead to increased social interactions, as these areas become focal points for community engagement and shared activities. This relationship is reviewed in [9], which examines how factors like social and community engagement, physical design, functionality, and sense of place contribute to vibrant public spaces. The study highlights that pedestrian-oriented, mixed-use neighborhoods are particularly effective in fostering social capital, underscoring the role of well-designed spaces in supporting strong community connections.

*Control:* While usage and social interactions contribute significantly to the vibrancy of public spaces, the element of control can deeply impact how freely people feel they can engage in these environments. A recent study introduces a tool for assessing the level of control in urban spaces by examining laws and rules, surveillance and policing, design and image, and access and territoriality, as explored in [68]. Findings indicate that privately owned public spaces typically impose more control over users, particularly through surveillance and restricted access, which can limit openness and alter the quality of social interactions within these spaces.

*Evaluation Using Indices and Indicators:* Evaluating public spaces through indices and indicators provides a structured, measurable approach to understanding quality factors from various perspectives, guiding improvements that align with users' needs. A set of indicators is particularly helpful as it offers clear criteria for assessing key aspects like accessibility, comfort, inclusiveness, and usability across different public spaces. This structured approach enables urban planners to identify strengths and areas needing improvement, allowing for targeted interventions that enhance the overall user experience. Additionally, indicators allow for consistent evaluations and comparisons between spaces, leading to well-informed design and policy decisions that meet community needs. To establish a framework for design evaluation, one study proposes a set of indicators for urban public spaces, focusing on six quality factors: publicness, identity, connectivity, spatiality, usability, and environment, as seen in [69]. This research emphasizes comparing priorities across stakeholder groups (practitioners, public officials, and academics) to ensure that diverse perspectives inform public space design and improvement.

Similarly, examining how specific public space indicators affect quality of life is essential, as it highlights how features like aesthetics, accessibility, and functionality influence residents' well-being, social engagement, and sense of place. In a distinct study, [8] explores three primary quality factors: aesthetic, semantic-perceptual, and activity-based functional, within a rural setting. The findings show a positive impact on rural vitality and dynamism, with physical and environmental indicators proving most influential. This underscores the value of tailored indicators for enhancing rural public space quality, illustrating that both urban and rural spaces benefit from targeted quality evaluations. This emphasis on tailored indicators is further explored in an urban context in a separate study from Gwangju, South Korea, which develops specific indices aimed at activating urban public spaces. This study, detailed in [70], examines factors such as environmental, human behavior, social, and physical elements to support dynamic urban engagement. Findings reveal that accessibility stands out as the most significant factor influencing park usage across diverse demographics, underscoring its essential role in fostering inclusive and adaptable urban spaces. While accessibility is highlighted as a crucial factor in supporting urban public space usage, other experiential qualities also play a vital role. [71] introduces an index that considers the user experience across four main quality factors: comfort, inclusiveness, diversity and vitality, and image and likeability. This study reveals notable differences in perceived quality among various public spaces, with aspects such as maintenance and inclusivity often scoring lower. Together, these studies underscore the multi-dimensional nature of public space quality, emphasizing that a blend of accessibility, comfort, and inclusivity contributes to the overall effectiveness of urban public environments. Expanding on the importance of experiential qualities in public spaces, a study in the new city of Ali Mendjeli in Constantine, Algeria, evaluates public space quality using the Public Space Index (PSI). As seen in [72], this research focuses on five main quality factors: inclusiveness, meaningful activities, comfort, safety, and pleasurability. Interestingly, despite the space being designed with minimal user input, it achieves a high-quality score, excelling in inclusiveness but showing limitations in comfort. This highlights that while inclusiveness can be effectively integrated into urban spaces, areas like comfort may require deeper user engagement and feedback for improvement.

*Evaluation using models:* In pursuit of a more comprehensive approach to assessing comfort specifically, [57] introduces a model-based assessment utilizing Building Information Modeling (BIM). This study evaluates five key quality factors, thermo-aeraulic comfort, visual comfort, acoustic comfort, wind comfort, and overall urban quality—through an innovative comfort rating system and an extensive framework with over 200 indicators. This model not only provides a detailed assessment of comfort but also offers a scalable method for evaluating public space quality, underscoring the potential of structured, model-based approaches to support adaptable and user-centered public space design. In addition to individual assessments, a model-based approach provides valuable tools for comparative analysis of public space quality across different contexts. [73] introduces a comprehensive framework specifically designed for town centers. This model evaluates six main quality factors: composition and enclosure, vitality and activities of needs, safety, and consistency with sustainable development ideas. By allowing for cross-town comparisons, this model facilitates a deeper understanding of how diverse factors like comfort, accessibility, and adaptability contribute to public space quality, supporting tailored improvements in town center environments.

Extending beyond comfort and standard quality dimensions, graphical analysis methods also offer unique insights into public space quality. [74] employs the RGBG Strategic Model to examine urban public spaces in Orihuela, Spain. This study assesses four key factors: grey (infrastructures and mobility), red (commercial and industrial fabric), blue (historic architecture, facilities, leisure, and tourism), and green (territorial green spaces, countryside, and water infrastructure).

The graphical analysis revealed various urban conflicts, particularly within the historic center, and led to targeted recommendations for enhancing mobility, commercial vibrancy, and the integration of natural elements. This approach highlights how graphical analysis can uncover underlying spatial conflicts and facilitate comprehensive strategies to improve urban public space quality.

While various models assess multiple quality factors, some focus specifically on individual dimensions. [75] introduces a targeted approach to evaluate the "publicness" of spaces. This model examines five main factors: ownership, control, civility, physical configuration, and animation. By applying these criteria, the Star Model enables a comparative analysis of publicness across spaces, highlighting areas where public qualities may be enhanced or, conversely, where they may be constrained. This focused assessment underscores the importance of publicness as a foundational dimension, providing urban planners with insights to foster more accessible, inviting, and community-centered public spaces. Beauty or attraction is also a factor that causes the public to visit a public space.

*Need for a comprehensive evaluation framework:* An evaluation framework is essential because it provides a structured approach to assess public spaces across diverse quality factors, offering insights into existing quality levels and identifying areas for improvement. Various studies emphasize the importance of these frameworks. For example, [76] introduces a Multi-Criteria Decision Analysis (MCDA) framework that demonstrates how spatially explicit assessments can guide urban quality evaluations. This study focuses on four quality factors: accessibility, livability, vitality, and identity. In support of a more spatially inclusive approach, [77] suggests that criteria related to architecture and urbanism are crucial for a comprehensive evaluation of urban spaces. This research analyses international tools used to assess urban life quality and identifies seven main quality factors: Architecture and urbanism, infrastructure, nature, health and well-being, social environment, development, and supplementary criteria. A similar approach is used in [78], where socio-spatial relations are examined through space syntax analysis, revealing correlations among spatial integration, visual exposure, and activity distribution. This study focuses on five main quality factors: spatial configuration, visual integration, functional distribution, physical qualities, and social qualities.

In a comparable line of research, [79] investigates spatial qualities, user interaction, and liveliness, presenting a computational approach to assessing public space quality. It considers four main quality factors: Liveliness, People, Interaction, and Space. The study finds that the descriptive attributes of neighbourhoods significantly influence liveliness, emphasizing the importance of spatial characteristics in creating engaging public areas. Liveliness itself is essential for making public spaces vibrant and welcoming, encouraging people to gather and interact. This is further illustrated in [54], which analyses how location, comfort, attendance, activities, and management contribute to public space quality. While Irkutsk's public spaces exhibit strong physical attributes, the study finds they fall short in promoting social inclusion and civic engagement, two aspects closely tied to liveliness and community connection. Expanding on the importance of these qualities, [80] assesses public spaces through economic, social, cultural, and urban perspectives. This study highlights those representative downtown squares, with their rich social and cultural attributes, most effectively fulfil public functions. Together, these studies underscore that vibrant, well-utilized public spaces contribute significantly to the social and cultural life of a community. Beyond the social and cultural qualities of public spaces, accessibility is another critical factor in making spaces inclusive and welcoming for all. [81] highlights accessibility alongside other essential qualities like safety, human scale, and interactivity, noting that while aspects like legibility and identity are well-developed, interactivity and flexibility are less emphasized in large urban settings. This finding emphasizes the need for adaptable, accessible spaces that meet diverse urban demands.

Supporting this perspective, [82] examines public space quality through the lens of sustainable development goals. It identifies accessibility, along with inclusiveness and design strategy, as crucial for creating urban projects that balance structural conditions with positive public use. Similarly, [83] analyses accessibility as a core quality factor in Iranian urban spaces, specifically in Hamadan. Findings indicate that limited proximity to key areas affects accessibility, while vitality is constrained by inadequate space for social gatherings and poor access to retail zones, diminishing the effectiveness of these urban spaces. Accessibility also intersects with issues of surveillance and inclusivity, as highlighted in [84]. This study critiques the extensive control and surveillance in Dubai's public spaces, which impacts accessibility and the cultural relevance of these areas. The findings suggest that while accessibility is crucial, its effectiveness can be limited by factors like user freedom and cultural expression.

In a similar analysis focusing on accessibility, functionality, and aesthetics, [85] reveals that accessibility ranks highest among quality factors. However, issues such as cleanliness and certain design elements like litter bins and advertising pillars detract from overall quality, underscoring the need for accessibility to be complemented by well-thought-out amenities that enhance urban space appeal and usability.

Assessing public space quality in rural and suburban contexts highlights distinct challenges and priorities compared to urban areas. A study on public open spaces in Engure, a rural coastal village in Latvia, evaluates quality, accessibility, and functionality as key factors, finding that the overall quality is moderate, with areas such as greenery and general upkeep needing improvement. This research underscores the importance of tailored approaches to meet the unique needs of rural communities, where specific environmental and functional aspects play a crucial role.

Examining both urban and rural settings, a comprehensive study on Polish public spaces identifies eight quality factors: functionality, practicality, reliability, durability, safety, legibility, aesthetics, and sensitivity [6]. Findings reveal notable differences in quality across contexts, with urban spaces displaying greater variability in quality levels. This study emphasizes that effective public space development must consider the diverse qualities required in different environments. Focusing on suburban public spaces, another study presents a methodology for evaluating recreational areas, identifying diversity, management, accessibility, vitality, integration, and activity as primary factors influencing space quality [86]. The research finds that diversity most significantly impacts the utility of recreational spaces, and that semi-public spaces are essential for fostering social life in suburban areas. This suggests that suburban recreational spaces benefit from a community-centered approach that values accessible and diverse activities.

In evaluating urban spaces, diverse approaches and frameworks reveal the importance of accessibility, social engagement, and adaptability in creating vibrant and inclusive environments. The variation in focus across studies, where some researchers examine certain quality factors while others address different ones, underscores the need for a comprehensive framework. Holistic assessment of public spaces can be challenging without a unified approach, as individual evaluations may overlook essential elements. To achieve a complete evaluation, public spaces should be divided by spatial type or context, allowing identification of quality factors specific to each setting. The following subsections provide an in-depth review of the quality factors relevant to assessing various public space spatial types.

## 4.2. Open spaces

The evaluation of open public spaces from a multidimensional perspective reveals varied approaches to assessing quality, each offering unique insights into the factors contributing to vibrant and functional urban environments. [87] lays the groundwork by presenting a comprehensive quality assessment directory for multi-functional public spaces, encompassing areas such as streets, squares, and parks. This directory identifies five key quality factors: inclusiveness, desirable activities, comfort, safety, and pleasure and organizes them into a detailed framework with 42 sub-criteria. By providing such a structured assessment, this directory serves as a foundational model that informs the understanding of complex quality dimensions in urban spaces. Similarly, [26] introduces a comprehensive model for assessing outdoor public spaces that focuses on six main quality factors: covering materials, environmental conditions, quality of life, social interaction enhancing, amenities and accessibility, and sustainable management. In addition to identifying critical factors, this model includes a relationship matrix between criteria and sub-criteria, highlighting specific elements that influence public space quality most. A similar approach is seen in [88], that employs a spatial multicriteria decision analysis approach to assess urban quality, emphasizing aspects such as physical setting, connectivity, vitality, meaning, and protection. This study introduces suitability maps (providing scores) that depict urban quality scores across various neighborhoods, allowing for a spatial visualization of quality differences within a city. This mapping approach complements the previous models by providing a geographical perspective on public space quality, which can help direct targeted interventions in areas identified as lacking in specific quality dimensions. [89] also does four-fold methodological approach data collection and analysis including semi-structured interviews, attitudinal questionnaires, observation study and space syntax analysis. The study showed that both the physical and the spatial qualities of the open space under study did not meet the user's needs and expectations. This includes aspects of cleanliness, maintenance, safety and the open space layout design and quality. On the other hand, accessibility has shown to achieve a satisfactory level.

While the previous papers on multidimensional analysis provide broad frameworks for evaluating public space quality, [90] offers a more user-focused perspective by assessing accessibility, comfort, uses, and sociability. The study's findings indicate that although public spaces are widely available, many users perceive their quality as lacking due to factors such as insufficient government management and lack of community involvement. This underscores the importance of governance and management practices in public space quality, emphasizing that structural assessments must be balanced with effective oversight to ensure that spaces meet community needs. In a similar way, [91] focuses on quality assessments specifically within town centers and stresses the significance of sustainable development principles. The study identifies six main quality factors, including accessibility, safety, vitality, and adaptability, and highlights that well-planned revitalization efforts are crucial for enhancing public space quality. This aligns with the governance focus of [90], suggesting that management and sustainability considerations play essential roles in the long-term success of public spaces, particularly within urban centers where revitalization can directly impact community engagement and accessibility. Furthermore, for the study done in [92] for the case of evacuation during disasters for people with disabilities, it emphasizes the need to avoid universal accessibility standards implementation shortcomings.

Complementing these perspectives, [21] explores the relationship between quality and sense of place in public spaces. The study finds that factors such as functional building aspects, visual elements, and urban experience significantly contribute to space's sense of space, while functional access has a lesser impact. This insight enriches the broader quality assessment by emphasizing the role of perceptual and experiential elements, highlighting that public spaces are not only defined by their physical structure but also by the emotional and sensory responses they evoke in users.

Following the foundation of comprehensive models for public space evaluation, the focus shifts to technology-based methods, offering precise and dynamic insights into quality dimensions essential for evolving urban needs. These methods have introduced diverse approaches to assessing public open spaces, capturing factors such as aesthetics, accessibility, user activity, and environmental quality.

On a more user-specific level, the [93] targets health-oriented aspects of public space usage by focusing on activities, comfort, and safety. Unlike image-based assessments, this app enables data collection on how spaces support physical activity, providing health researchers and urban planners with practical tools to understand public space functionality in promoting well-being. Finally, the [94] offers a high-level model to predict potential user presence and movement within different urban configurations. By evaluating factors like accessibility, attractiveness, movement patterns, and temporal variations, it provides insights into how spatial designs influence user interactions. This model adds a predictive element to public space assessment, offering city planners a way to forecast and design spaces that encourage user engagement based on urban form. Together, these studies form a layered approach to evaluating public spaces, integrating aesthetic, functional, behavioral, and predictive dimensions.

To understand public space quality, it is crucial to evaluate specific factors that impact safety, security, mental health, and user experience. [95] introduces an indicator system for urban space safety, identifying lighting, pedestrian traffic, and police presence as the most influential factors for enhancing street safety and coexistence. Similarly, [96] supports the importance of lighting, demonstrating that higher levels of illumination and uniformity contribute positively to pedestrians' sense of safety, with warmer lighting further enhancing this perception. These findings underscore how lighting and surveillance are foundational to the perception of safety in public spaces. [97] adds to these findings by evaluating public spaces from the perspective of elderly users and identifying safety as a prominent concern alongside accessibility, comfort, and usability. This study highlights gaps in design and management that affect safety for elderly users, and issues with pavements and pollution. Addressing the control and publicness of public spaces, [98] examines how these factors shape user behavior and engagement through a spatial model. By evaluating ownership, animation, civility, and physical configuration, the study reveals how different mechanisms of control impact openness and safety in parks, streets, and squares. Control is shown to be essential in balancing accessibility with security, contributing to a comprehensive understanding of public space quality.

[99] extends the understanding of security and comfort by examining how perceptions of public space vary among residents in displaced relocation settings. This study identifies comfort, security, and pleasure as key factors and reveals that perceptions differ notably before and after relocation. This highlights the importance of community integration and consistent public space quality to support positive user experiences across different demographic settings. The link between the perception of safety, security, and mental health reinforces the idea that these aspects of public spaces are

interdependent and shape users' experiences. Utilization emerges as another critical factor that contributes to overall public space quality. [100] compares open space utilization across three Asian cities, finding significant variations in usage due to local context, culture, social values, and climate. This illustrates how public space utilization is dynamic and context-specific, with effective design and accessibility being key to encouraging active use. Similarly, [101] examines utilization alongside physical features in open spaces, emphasizing the importance of comprehensive park design that addresses multiple qualities, such as safety, comfort, and attractiveness. Inclusive features, such as walkable areas, are crucial to encourage usage, especially when considering diverse user needs.

Expanding on the theme of accessibility, [102] explores the importance of walkable, green public spaces through a web-based tool called PedestrianCatch. This tool evaluates accessibility, inclusivity, and walkability by simulating pedestrian catchments and accounting for connectivity, topography, and walking speed. The study reveals that distributed greenery and well-designed open spaces significantly enhance accessibility and thermal comfort. This finding underscores the impact of thoughtful, inclusive design on both physical and psychological comfort, reinforcing the interconnectedness of accessibility, safety, and overall public space quality.

Together, these studies illustrate the multifaceted nature of open public space evaluation, highlighting the importance of safety, security, mental health, utilization, control, accessibility, and comfort. Examining each of these factors, individually and in relation to one another, offers a deeper understanding of what makes public spaces effective, inclusive, and supportive of diverse community needs. Among these factors, comfort stands out as a consistently studied quality, encompassing thermal, visual, and acoustic aspects that are essential for enhancing user satisfaction and shaping positive interactions within public spaces.

The comfort of public spaces plays a critical role in addressing climate change challenges in urban areas. In [103], the study examined five key quality factors (QF): comfort, safety and security, accessibility, legibility, and inspiration and attractiveness and found that street orientation is a crucial determinant of thermal comfort. By optimizing street design, particularly with distributed greenery along East-West axes, public spaces can effectively manage thermal environments, thus contributing to overall climate resilience. Comfort in public spaces is often understood as a composite of multiple sensory factors, including auditory, visual, and thermal comfort [104]. Among these, auditory comfort has the most substantial impact on overall user comfort, highlighting the importance of managing soundscapes in urban areas. [105] extends this perspective by studying the thermoacoustic environment and its effects on subjective evaluations of comfort. This research emphasizes that both acoustic and thermal factors are critical, although thermal factors exert a greater influence on perceived comfort in outdoor spaces. [106] explores the variability of thermal environments across urban settings, confirming that elements such as water bodies, vegetation, and urban morphology significantly contribute to thermal comfort. This finding aligns with the earlier observations in [103], which identified greenery and urban design features as essential components of climate-sensitive urban planning. However, [107] offers a contrasting yet supporting view, suggesting that while vegetation and urban geometry enhance thermal comfort, the presence of a water body alone does not significantly improve comfort or urban livability. Instead, [107] links thermal comfort directly to broader measures of urban livability, emphasizing the need for an integrated approach to public space design. Thermal discomfort is particularly pronounced in street environments. [108] found that streets have the worst thermal comfort conditions compared to other urban spaces. Supporting this observation, [109] underscores the importance of street orientation in determining overall comfort, reinforcing the critical role that urban planners must play in optimizing street layouts to mitigate discomfort. Further elaborating on the relationship between urban design and thermal comfort, [110] introduces the height-to-width ratio of streets as a crucial factor in enhancing comfort. This study suggests that compact urban designs, characterized by optimal street dimensions, can help control microclimates and improve thermal conditions, thus ensuring that public spaces remain comfortable for users even in extreme weather conditions. Another interesting study in [111], Evaluation of the thermal environment in an outdoor pedestrian space.txt, evaluated the thermal environment in urban pedestrian spaces using two quality factors: thermal environment and subjective thermal sensation. The study found that thermal sensations correlate with measured air temperatures but are also significantly influenced by factors such as solar radiation, wind velocity, and the pedestrian's recent thermal history. A much-related study with similar findings in [112], Perception of temperature and wind by users of public outdoor spaces relationships with weather parameters and personal characteristics.txt, investigated the perception of temperature and wind in public outdoor spaces. It focused on three main quality factors: thermal environment, wind environment, and personal characteristics. The study found that thermal comfort in outdoor spaces is influenced by a combination of environmental factors and personal characteristics, with a comfort range for air temperature between 23 and 28°C.

The concept of acoustic comfort in public spaces has expanded with the introduction of new dimensions to better capture how soundscapes affect user experiences. [113] introduces the soundscape dimension with four key quality factors: relaxation, communication, spatiality, and dynamics. The study found that the perceived dominance of sound sources, especially natural sounds, had a significant impact on these dimensions, with natural sounds enhancing relaxation levels. This suggests that incorporating natural soundscapes into urban design can significantly improve acoustic comfort. [114] looks at soundscape quality and overall environmental quality, showing how different types of urban spaces, whether parks, cultural areas, or pedestrian zones, require tailored strategies to improve acoustic environments. The study reinforces the idea that acoustic comfort is highly context-dependent, and improvement strategies must reflect the unique characteristics of each space.

Comfort in urban open spaces is shaped by both thermal and acoustic factors, with aspects like street orientation, vegetation, and soundscapes playing pivotal roles. Creating spaces that offer both thermal and acoustic comfort involves context-sensitive designs, incorporating natural features and sound management strategies to improve the overall experience for users. In assessing open spaces, a wide range of factors come into play. While evaluating large areas, such as entire cities, can be challenging, it is often more effective to assess these spaces based on specific types, such as parks, squares, streets, or even narrower categories like pedestrian zones or water parks. [115] presents an innovative approach to analysing public open spaces, using techniques like urban morphology, parametric modelling, and data mining. This study focuses on eight key quality factors: void shape, vertical plane and permeability, urban indices and density, visibility and connectivity, urban system, use and appropriation, environment, and generic labels. This approach shows strong potential for creating more detailed classifications of public spaces and supporting urban design grounded in data.

Using similar methodologies to assess even smaller spaces can lead to a comprehensive understanding of public spaces at various scales. The goal is to identify a full range of quality factors so that multidimensional assessments reflect all relevant aspects. The following sections review different types of open public spaces and the specific quality factors that contribute to their assessment, supporting a more accurate evaluation of these diverse environment.

*4.3. Green spaces*

Urban green spaces play a crucial role in enhancing the quality of life in cities, offering a wide range of environmental, social, and health benefits. Several studies have focused on the assessment of these spaces, highlighting key quality factors such as accessibility, attractiveness, and user perception. To understand the complex nature of urban green spaces, researchers have developed various evaluation frameworks and identified key quality factors. A significant contribution to this field is made by the research in [116] that presents a comprehensive evaluation framework that considers five main quality factors: accessibility, aesthetics, affordability, spaciousness, and surface quality. By identifying significant disparities in green space provision and quality across different areas, the study emphasizes the necessity of integrating both accessibility and quality into urban green space assessments to ensure equitable distribution and enhanced user experience.

Two studies [117], [118], specifically emphasize the importance of accessibility and attractiveness in evaluating green spaces. The first study adopts a functional approach, assessing a variety of green space types, including parks, forests, nature reserves, and farmlands. It uncovers deficiencies in green space provision at different functional levels and examines the impact of barriers on accessibility, offering insights into how these limitations can be addressed to enhance urban green environments. Similarly, [118] employs a GIS-based model to assess green spaces, focusing on proximity (accessibility) and quality (both inherent and use-related). Concentrating on parks, forests, and nature reserves, this study highlights spatial disparities in green space distribution, revealing a pattern where central urban areas exhibit lower accessibility and quality compared to peripheral regions. Both studies emphasize the need to improve accessibility from a functional and distributional point of view.

Expanding the scope of assessment, [119] introduces a more comprehensive approach by evaluating parks based on ecological, microclimatic, and social quality factors. This study found significant differences between natural and artificial varieties of structural elements across all dimensions, with natural varieties scoring higher in ecological and social dimensions. In contrast to the previous studies, which primarily focus on accessibility and internal qualities, this paper emphasizes the structural aspects that contribute to the overall quality of green spaces. Similarly, [120] employs a trivalent framework to evaluate ecological, socio-cultural, and aesthetic qualities of public green open spaces. This study highlights that many public open spaces are not optimally planned, designed, or managed, underscoring the necessity of considering multiple dimensions in public space planning. Together, these studies advocate for a more holistic evaluation of urban green spaces, integrating structural and functional quality factors to enhance their overall effectiveness.

While these evaluation frameworks provide valuable insights into the structural and functional aspects of urban green spaces, recent research has increasingly focused on another crucial dimension: the user experience, with user perception and comfort emerging as critical themes in understanding how people interact with and creating truly effective urban environments. For example, [121] specifically examines thermal comfort in vegetated areas and its relationship to visit intensity. The study found that most visitors feel uncomfortable or very uncomfortable in space, attributing this discomfort to the low percentage of vegetated areas and a high thermal humidity index.

This research emphasizes the importance of proper design and vegetation in urban green spaces to enhance thermal comfort for users. Complementary this, [122] broadens the perspective on thermal comfort by incorporating additional factors such as nature and biodiversity, quietness, historical and cultural value, spaciousness, cleanliness and maintenance, facilities, and the feeling of safety. The study found that cleanliness, maintenance, quietness, and safety were perceived as the most important qualities by users, while naturalness and historical value were rated lower. This highlights the multifaceted nature of user perceptions in urban green spaces, suggesting that while thermal comfort is essential, other quality factors also significantly influence the overall user experience.

Both the [123] and the [124] contribute significantly to enhancing user experiences in urban environments by focusing on the quality of green spaces, yet they differ in scope and focus. The community-scale assessment adopts a broader perspective, examining public green spaces within neighborhoods and their functionality. It identifies eight key quality factors, including vegetation configuration, water condition, and road condition, revealing how transportation modes can create spatial imbalances that exacerbate inequities in access to quality green environments. In contrast, the roadside study specifically zeroes in on the aesthetic and maintenance aspects of greenery along roadways, evaluating visual quality through factors such as biophysical structure, naturalness, and maintenance. While the community-scale study addresses the broader implications of green space accessibility, the roadside assessment emphasizes the importance of visual appeal and upkeep in enhancing user satisfaction.

While user perception is indeed critical, research has demonstrated that visual perception encompasses a much broader range of elements. [125] focuses on users' visual perceptions of urban vegetation landscapes. It examines elements such as vegetation structure, height ratio, exfoliation appearance, planting density, color contrast, and species diversity. The research found that strong color contrast and the combined use of evergreen and deciduous plants enhance visual aesthetic quality. Thus, beyond functional and environmental factors, the visual design of green spaces plays a critical role in shaping user experiences. These user centric studies collectively underscore its significance in urban green space assessment, revealing that a comprehensive understanding of various quality factors including thermal comfort, cleanliness, safety, and visual aesthetics is essential for enhancing visitor experiences. Together, these studies illustrate the interconnectedness of quality factors of green space types and underscore the necessity for a comprehensive approach with an emphasis on functional accessibility and visual quality.

### 4.4. Parks and Waterfronts

Parks are of immense value to urban environments, not only for their recreational benefits but also for their economic, social, and ecological contributions. [126] estimates the value of urban parks, focusing on three key quality factors: economic value, social value, and ecological value. The study found that the recreational benefits provided by urban parks were comparable to the city's annual expenditure on maintaining green spaces, highlighting the importance of considering the full benefits of parks in city planning and management. This underscores the need for proper evaluation and maintenance to preserve their quality and ensure long-term benefits.

In evaluating public parks, several studies have employed multidimensional quality assessment indices, each designed to capture specific aspects of park utilization and user experience. For instance, [127] introduces the Space Utilization Index (SUI) to assess a public park in Indonesia. This study examines four key quality factors: space utilization, user diversity, time utilization, and activity diversity. A high SUI score signifies effective utilization, reflecting how well the park accommodates a diverse range of users and activities throughout the day, highlighting its functional adaptability. In contrast to the focus on activity and user diversity throughout the day, [128] emphasizes structural play diversity, particularly from the perspective of youth using QUINPY index.

The study also considers nature, park size, maintenance, and safety as critical dimensions. This evaluation highlights the importance of diverse play structures and natural elements, which are crucial for engaging younger park users and ensuring their safety and enjoyment. Expanding the scope, [129] examines public open spaces in India using the Public Open Space Index (POSI) for a comprehensive evaluation. This index integrates five quality factors: individual well-being, inclusiveness, engagement, sustainable spaces, and management. The study stresses the role of sustainability, showing how environmentally and socially inclusive parks can foster user engagement and contribute to well-being, positioning parks as integral components of urban planning strategies. Similarly, [130], using the Good Public Space Index (GPSI), also focuses on inclusiveness and meaningful activities but introduces other key factors such as comfort, safety, and pleasurability. While both studies highlight the importance of inclusiveness, [130] additionally underscores the challenges of ensuring safety and comfort in high-traffic areas, pointing to the need for improved design and management in densely populated urban spaces. Building on this, [131] extends the GPSI framework to assess mid-sized city parks, examining their applicability to smaller urban settings. The research revealed wide variation in park quality, particularly in maintaining meaningful activities and pleasurability, which proved more challenging in smaller parks.

These findings underscore the difficulties of applying an index originally developed for larger urban spaces to mid-sized environments, where space limitations may impact activity diversity and overall enjoyment. This highlights the need for adaptable evaluation tools that consider the specific constraints and dynamics of smaller parks. Since safety is identified as a major quality factor by the public, the evaluation of safety is necessarily important. [132], safety evaluation methodology of urban public parks by multi-criteria decision making.txt, proposes a methodology for evaluating safety in urban public parks. It focuses on nine main quality factors including surveillance, sightlines, lighting, perimeter control, access control, activity support, maintenance, surrounding neighborhood socio-economic context, and crime. The research found that surrounding socio-economic context has a major impact on park safety. Unlike the previous five studies that employed indices for park evaluation, the following studies focus solely on quality factors to assess park performance. For instance, [133] assesses public spaces in Bangkok using a multidimensional set of quality factors, including accessibility, imageability, usability, and sociability. These factors capture the broad ways in which parks are experienced, from how easily they can be accessed to their visual and social appeal. The study reveals significant differences in park quality across urban zones, with inner-city parks scoring higher across most dimensions, suggesting an urban divide in the quality of public space provision.

Supporting this observation, [134], which evaluates both the quantity and quality of parks in Hail City, also found considerable variation in park quality across the city. Using eight quality factors (surrounding environment and accessibility, quietness, internal environment, water features, services and facilities, health issues, security and safety, and aesthetics), the study revealed a significant shortage of park space, further highlighting the disparities in park provision and quality. Similarly, [135] evaluates parks through quality factors but shifts their focus to more practical aspects such as maintenance and growth. It considers quality, sufficiency, maintenance, and growth rate, with a key emphasis on user feedback as the primary source of evaluation. This transition from broader environmental factors to user-centric insights highlights the role of public perception in assessing practical park attributes like upkeep and sufficiency, which directly impact overall park satisfaction. Likewise, [136] values user input but takes a more architectural and functional approach. It evaluates three key quality factors: architectural quality, functional quality, and contextual quality. While still relying on user assessments, this study emphasizes how the physical arrangement and maintenance of park elements affect perceived environmental quality. Additionally, the research explores the restorative capacity of parks, showing how functional and architectural order can enhance user well-being.

In line with the multidimensional evaluation seen in earlier studies, [137] employs a comprehensive set of indicators to assess park quality, focusing on three key quality factors: natural elements, built elements, and spatial context. The study highlights that smaller parks often score higher in terms of quality, particularly when context measures were considered. This suggests that even smaller parks can offer high-quality experiences when their surrounding environment and design elements are carefully integrated, demonstrating the importance of spatial context in enhancing park quality. Taking a different approach, [138] evaluates park quality using the space quality diagram, focusing on four key quality factors: access & linkages, uses & activities, sociability, and comfort & image. Conducted for a park in Türkiye, this method provides an overall quality value while also identifying specific areas in need of improvement, particularly related to maintenance issues. The space quality diagram offers a holistic assessment, enabling targeted interventions to enhance park quality and management by highlighting both strengths and areas requiring attention. In contrast, [139] employs a quantitative approach, using three main quality factors: needs, rights, and meanings. This method could be applied to evaluate whether a park has met established standards, providing a structured means to assess whether essential aspects of a park's quality are being fulfilled.

More focused assessments, such as green space quality evaluation within a park, should be conducted as these spaces often play a key role in community well-being, providing environmental, social, and recreational benefits that must be maintained to ensure the park serves its full purpose. [140] investigates the determinant factors for quality green open space assessment, particularly in neighborhood parks. It identifies eight key quality factors: social sustainability, use pattern, place attachment, facilities, safety and security, nature preferences, activities, and accessibility. Among these, social sustainability emerged as the most significant factor, highlighting the importance of fostering long-term social cohesion and engagement in the design and management of neighborhood green spaces. User preferences and perceptions are essential in this process, as they help evaluate the success of public spaces like parks by ensuring these spaces are effectively tailored to meet the community's needs. [141] analyses tourists' preferences for public space in Fuzhou national forest park using big data from online comments. This study examines six main quality factors: emotional response, activities and attractions, environmental features, temporal aspects, accessibility and transportation, and management and facilities. The findings revealed overwhelmingly positive emotional responses, popular attractions, and scenic spots, but also highlighted areas for improvement, particularly in crowd management during peak times.

Similarly, [142] shifts the focus from tourist preferences to users' perceptions of inclusiveness in urban parks. While both studies highlight the role of emotional responses and environmental features, [142] further investigates how physical characteristics and individual experiences shape inclusiveness. It identifies eight key quality factors: site conditions, site usage, independent mobility, supply-demand matching capacity of space and facilities, interference and limitation of recreational activities, positive emotional experience, broad social participation, and place identity. The study found that inclusiveness is influenced by both environmental factors and personal emotions, emphasizing the importance of creating spaces that are not only physically accessible but emotionally welcoming. In a related way, [143] explores how park users perceive the health and well-being outcomes associated with urban green spaces. [144] evaluates the inclusivity of thematic parks compared to non-thematic parks. It focuses on four main quality factors: physical access, social access, access to activities, and user characteristics. The research found that while non-thematic parks scored slightly higher in overall inclusivity, thematic parks had higher visit frequency. Inclusiveness is indirectly related to well-being. [145] does the study on well-being, considering four key quality factors: physical environment, social interaction, health and well-being, and park management. The study found that over 80% of respondents reported positive effects on stress reduction, mindfulness, physical fitness, and overall well-being from park usage, reinforcing the significant role parks play in enhancing both mental and physical health.

Comfort is a crucial aspect of user perception, shaping how public spaces like parks are experienced. Expanding the focus on comfort to thermal conditions, [146] explores user perceptions of thermal comfort in an outdoor urban park in Denpasar. The study revealed a gap between users perceived thermal comfort and objective measurements, with users finding the park more comfortable than simulations suggested. This highlights the importance of considering subjective user perceptions in addition to objective data when assessing comfort in outdoor spaces. In contrast to the subjective perceptions, it focuses more on the meteorological and landscape factors affecting outdoor thermal comfort in urban parks. It identifies three main quality factors: meteorological factors, landscape types, and thermal indices, finding that air temperature was the most significant factor impacting comfort across different landscape types and seasons. This objective analysis complements the user-centered findings of the other studies, showing how both physical and perceptual factors contribute to overall comfort in public spaces. These studies show that comfort in public spaces depends on both the physical environment and how users perceive it.

Soundscape evaluation is also a crucial aspect of how users experience public spaces, particularly in parks where environmental sound quality can influence both individual and group activities. [147] investigates how environmental sound quality impacts visitors' soundscape perceptions and preferences in urban parks. It identifies four main quality factors: acoustic environment characteristics, soundscape perceptions, soundscape preferences, and visitor behaviour. The study found that visitors' perceptions of loudness and satisfaction were more closely related to maximum sound levels rather than average noise levels, highlighting how peak noise can affect overall user enjoyment in public spaces. Similarly, [148] also explores soundscape evaluations but focuses on how users' activities influence their perceptions of sound in an urban setting. While [147] emphasizes general preferences for environmental sound quality, [148] examines how different activities impact users' satisfaction with the soundscape. It identifies three key quality factors: disruption, stimulation, and suitability, revealing that socially interactive users rated their soundscapes more positively than solitary users. This contrast between general sound preferences and activity-specific perceptions underscores the diverse ways in which soundscapes are experienced depending on user activities.

While comfort and soundscape evaluation play important roles in shaping user experiences, a more comprehensive environmental assessment can provide a deeper understanding of urban park quality. [149] presents a methodological approach to the environmental quantitative assessment of urban parks. It compares the environmental performance of parks, urban squares, and street canyons, focusing on three key quality factors: thermal stress, air pollution stress, and noise stress. The results indicate that urban parks demonstrate superior environmental quality compared to other urban spaces in both summer and winter, due to the moderating effects of vegetation on thermal conditions and air pollution levels. Similarly, [150] expands on environmental quality by analyzing public green spaces in Bartın, Türkiye, through a sustainable urban fabric lens. This study identifies five main quality factors: functional quality, aesthetic quality, technical quality, ecological quality, and economic quality. Unlike [149], which focuses on the direct environmental stressors, [150] considers broader factors such as ecological and economic sustainability, revealing variations in the quality and usage patterns of different green spaces. This highlights the importance of both immediate environmental stress mitigation and long-term sustainable planning to enhance the overall quality of urban green spaces.

Waterfronts represent a distinct category of urban parks, offering green spaces that not only provide recreational and social value but also integrate environmental and aesthetic qualities due to their proximity to water. As with other types of parks, evaluating waterfronts requires a comprehensive approach that considers multiple dimensions ranging from accessibility and functionality to design and environmental sustainability. A quality factor specific evaluation is carried out in [149], which focuses on service capacity and accessibility in waterfront parks. Through a combination of Point-of-Interest (POI) data and user surveys, the study highlights how users prioritize waterfronts that are easily accessible preferably within a 15-minute walk or bike ride while also valuing high-quality visual connections to the surrounding landscape.

When considering the physical and aesthetic aspects, [151] shifts the focus to user perceptions of the visual and spatial qualities of waterfront spaces. The study reveals that harmony in design where physical elements blend seamlessly with the natural environment has the strongest positive effect on user preferences. In contrast, excessive spaciousness and lack of visual diversity detract from user experiences. This aligns with the design and aesthetic factors outlined in [152] where successful waterfronts are those that harmonize natural and built environments. These insights underscore the importance of creating aesthetically pleasing waterfronts that balance natural beauty with functional space for users [151]. Focusing on the social dimensions, [153] examines how publicness and inclusiveness play pivotal roles in the evaluation of waterfront parks. The research highlights that restrictions in access or strict rules can hinder inclusiveness, limiting the ability of waterfront parks to serve diverse populations. Public access and inclusivity are essential for fostering social interactions and ensuring that waterfronts enhance public life [129]. [154] presents a tool for assessing urban blue spaces considering social, aesthetic, physical, and environmental quality factors. The study developed a structured assessment tool covering these aspects, with potential applications in site comparisons, intervention assessments, and ongoing monitoring. This highlighted the fact that environmental aspects should also be considered for the assessment of a waterfront public space. While [155] evaluated the effectiveness of a waterfront public open space using three quality factors: responsive, democratic, and meaningful, using the Good Public Space Index (GPSI), the study found high overall effectiveness, with strengths in social interactions and diverse activities, but identified challenges in daytime usage and facilities for children and elderly visitors.

Thus, the evaluation of waterfronts, as reflected in these studies, shows a clear interconnectedness between service capacity, physical design, and social inclusiveness. The comprehensive framework from [156] serves as a useful guide for assessing how different quality factors come together to influence user satisfaction and the overall success of waterfront parks. However, as seen in the case of the Maltepe Fill Area, there are still gaps in urban integration and functionality that need to be addressed. From multidimensional approaches to specialized assessments, and from user preferences to environmental quality evaluations, these studies highlight the diverse factors essential for a comprehensive evaluation of both parks and waterfronts. Waterfronts, as a distinct subset of urban parks, require a careful balance between urban integration, accessibility, physical design, and social inclusiveness. To ensure that both parks and waterfronts are well-functioning, inclusive, and sustainable, it is crucial to develop a unified evaluation system that integrates all relevant quality factors, ranging from service capacity and aesthetic design to public access and environmental sustainability. Such an approach will allow urban planners to prioritize and address key areas of improvement, creating public spaces that truly enhance the quality of life for all users.

*4.5. Streets and Squares*

The evaluation of public spaces, particularly streets, involves understanding the interplay of multiple quality factors. [157] provides a key starting point by emphasizing the relationships between different dimensions of public spaces specifically the physical, activity, social, and meaning dimensions. The study found that the strongest correlation exists between the social and meaning dimensions, highlighting how life and interactions in a city are deeply tied to the meaning people attach to space. Social dimension encompasses the vibrancy and interactions within urban environments, often referred to as the "life" of a city. This concept is explored in [158], which examines the quality of public space in an urban village, focusing on accessibility, protected property, comfort, and enjoyment as key quality factors. Similarly, [159] discusses life in public spaces, identifying access and linkage, comfort, and image, uses and activities, and sociability as essential factors. The main finding of this study is that sidewalks and connectivity contribute significantly to vibrant urban life, reinforcing the importance of infrastructure in enhancing social interactions. Previous studies, such as [159], highlight how sidewalks and connectivity are crucial in fostering vibrant urban life. Extending this idea, [160] introduces the broader concept of walkability, which is defined by key elements such as density, mix, and access that shape how people move and interact within urban environments. This notion is further explored in [161], which develops a Walkability Assessment Tool (WAT) and applies it to historic districts, focusing on functional, emotional, and social affordances that impact walkability. Narrowing the focus, [162] identifies specific criteria for assessing pedestrian spaces, including factors like design, maintenance, activities, accessibility, and safety. Similarly, [163] evaluates pedestrianized streets, emphasizing how accessibility, safety, and comfort contributes to the overall quality and usability of these spaces, ensuring they support both walkability and urban life.

A similar concept is seen in [43], where a Global Walkability Index is introduced to assess pedestrian environments worldwide. It considered three quality factors: safety and security, convenience and attractiveness, and policy support.

The study developed a simplified index with 14 variables and created extended survey materials for detailed assessments. The research emphasizes the importance of local implementation and provides a framework for cities to identify specific areas for improvement in their pedestrian environments. However, public space quality extends beyond pedestrian environments to include broader street spaces. [164] bridges the gap between pedestrian spaces and the larger street environment, emphasizing the importance of pedestrian infrastructure in enhancing overall street quality. [165] offers a more general evaluation using multi-source data, covering broader street space quality.

In assessing public spaces, advanced images and simulation techniques have proven highly valuable. [166] employs quantitative methods for street quality evaluation, using deep learning and street view images to assess cleanliness, comfort, and traffic. [167] applies a more focused lens on specific street features, such as street vending activity. As these studies show, visual perception has emerged as a critical factor in public space evaluation. [168] proposes a methodology for measuring visual quality, considering both physical visual quality and perceived visual quality to provide a comprehensive assessment. Among the key quality factors for public spaces, safety stands out as a primary concern. It explores the perceived safety of street environments, while [169] extends this by examining the impact of improved street lighting on crime reduction, revealing that better lighting significantly reduces both crime incidents and the fear of crime, thus enhancing the safety and usage of streets after dark. A study in [170] also identified issues such as lack of continuity, poor attachment to surrounding urban tissues, insufficient open spaces for social interaction, and inadequate street lighting by studying a traditional commercial street in Belgium. It focused on two main quality factors: street quality and neighborhood communication. In addition to safety, several other quality factors play a significant role in determining public space quality. [103] focuses on liveability, evaluating spaces based on factors such as safety, convenience, legibility, comfort, inspiration, and liveability. This shows a transition from safety into broader quality concerns, illustrating the dependency between safety and other factors such as comfort and accessibility. Similarly, [171] examines the impact of street features on inclusiveness, revealing how certain street design elements can increase inclusiveness for a broader range of users. Meanwhile, [172] proposes a method to assess permeability, focusing on factors like frontage permeability, density, plot size, and network centrality to evaluate how public and private spaces interact.

Squares, as a part of streets in general [173], serve as vital urban spaces where people meet, interact, and experience the city. They are crucial elements of the urban fabric, offering a place for social interaction and various activities. Unlike streets, which primarily focus on movement and transit, squares are designed to encourage gathering and lingering, making them an integral part of public life. The evaluation of these spaces often revolves around their publicness, as highlighted in [66], which presents a framework for assessing urban public spaces but focuses specifically on the dimension of publicness, how accessible, open, and inviting these areas are to diverse groups of people. Maintaining footfall in these areas is strongly influenced by physical and environmental factors. One of the key factors in ensuring people use and enjoy these spaces is thermal comfort. [164] examines thermal comfort in public squares, specifically in hot, arid climates. It shows how different physical characteristics, such as vegetation density, surface coatings, and shading devices, contribute to creating distinct thermal environments within urban squares. The presence of these features plays a crucial role in keeping public squares usable throughout the year, particularly in climates where extreme temperatures are common. In addition to thermal comfort, acoustic comfort plays an equally significant role in shaping urban public spaces' quality. [161] delves into the concept of soundscapes and their impact on acoustic comfort in urban squares. It identifies factors such as relaxation, communication, spatiality, and dynamics as essential elements in evaluating acoustic comfort. The study highlights the positive correlation between acoustic comfort and these factors, noting that diverse types of sound (natural, human, or mechanical) affect the perception of comfort in varying ways, depending on how they align with users' activities and expectations.

User experience in these environmental factors is also critical of public space evaluation. As urban spaces are designed not only to be functional but also enjoyable, it is essential to understand how people perceive comfort in these environments. [170] examines the perception of comfort but with an additional focus on the quality of meaning in urban spaces. The study reveals how various aspects, such as physical and social dimensions, contribute to users' perceptions of public spaces. Beyond physical comfort, this study explores how users assign meaning to spaces based on their experiences, cultural associations, and social interactions, making it a valuable addition to understanding user experiences in public squares. The evaluation of potential squares using a quantitative method by considering identity as one of the factors along with other factors such as accessibility, size, characteristics of building, density, morphology, landscape character is sone in [174]. The research demonstrated varying potential among the studied squares, with some excelling in historical significance while others show strengths in accessibility or spatial quality. [175], evaluated parks and squares together in a city as a square could potentially come in a park. This paper evaluated the vitality of parks and squares in medium-sized cities. It focused on four main quality factors: urban functional area characteristics, travel vitality index of urban residents, park and square attractiveness, and regional service levels of parks and squares. The study found that functional mixing, road density, and population distribution significantly impact park/square vitality.

The literature reveals numerous quality factors discussed across numerous studies, yet street evaluation lacks a truly comprehensive approach as these factors continue to evolve. A systematic framework with well-defined and prioritized quality factors is crucial for accurate street space evaluation, enabling urban planners to address critical elements such as functionality, walkability, and inclusiveness, ensuring streets meet diverse user needs while remaining adaptable. Through strategic prioritization of these factors, cities can develop more vibrant, resilient public spaces that effectively support sustainable urban growth and enhance community well-being, transforming street evaluation from a fragmented process into a cohesive system that addresses both current needs and future adaptability.

### 4.6. Public Facilities

Public facilities are vital parts of any community, providing spaces for people to gather, learn, relax, and access essential services. They include schools, hospitals, libraries, playgrounds, courtyards, and transportation hubs. On university campuses, these facilities cover buildings, pathways, and outdoor areas that support learning and student life. One study examines the design quality of campus facilities using standardized Design Quality Indicators (DQIs), focusing on three primary areas: the indoor environment, safety, and maintenance; furniture, utilities, and space layout; as well as privacy, appearance, and the surrounding areas [176]. The research reveals areas of dissatisfaction, such as noise levels, natural light, and limited control over ventilation, suggesting a need for improvements to enhance user comfort and functionality. Another study takes a broader view, developing a model to evaluate quality of life across university campuses. This model considers five key factors: environment, mobility and parking, safety, urban space, and support services [177].

The model effectively captures differences in quality-of-life perceptions among campus users and evaluates how proposed changes could impact these factors, offering a comprehensive approach to campus facility assessment.

Focusing specifically on campus pathways, a third study investigates the visual, perceptual, and social dimensions of these spaces. It highlights the importance of diverse visual elements, the influence of architectural styles on user perceptions, and the potential for design to enhance social interactions along pathways [178]. This research underscores the value of pathways in shaping both the aesthetic and social experience on campus. Recreational spaces and playgrounds are also integral to public facilities, and their quality must be assessed to support community well-being. A study on public recreational spaces in Rio de Janeiro, Brazil, examines physical characteristics alongside the socioeconomic status of neighborhoods [179]. The findings reveal that public spaces in neighborhoods with lower socioeconomic indicators tend to have lower quality scores, highlighting the deprivation amplification phenomenon, where areas with fewer resources often experience reduced access to high-quality public facilities.

A study in [180] is also focusing on playgrounds to evaluate the effectiveness of non-green public spaces in a fisher settlement by considering three quality factors: democratic, meaningful, and responsive. The research employed the Good Public Space Index (GPSI) to assess six variables: intensity of use, intensity of social use, people's duration of stay, temporal diversity of use, variety of use, and Variety of Users. The study found that while the overall GPSI value was standard, there were strengths in social use intensity and variety of use, but weaknesses in accessibility and comfort factors. Research in Edmonton, Canada on playgrounds evaluates spatial accessibility and equity, focusing on five quality factors: spatial accessibility, spatial equity, playground quality, demographic need, and social need [181]. This study finds that playgrounds are equitably distributed, with greater accessibility observed in areas of high social need, indicating that equity considerations play a vital role in the distribution and quality of these recreational spaces. Expanding the focus on accessibility, a study presents a universal methodology for evaluating accessibility across diverse types of built environments, encompassing spaces such as playgrounds and public buildings [182]. This study emphasizes accessibility as the primary quality factor and reveals varying levels of accessibility, with architectural details and interior elements often showing the lowest accessibility scores. These findings highlight the importance of inclusive design across all aspects of the built environment.

Another type of built environment, courtyards, is studied in terms of quality through a distinct set of factors [183]. This research examines public courtyard quality with a framework that includes visual & morphological, functional, social, perceptual & experiential, and ecological dimensions. The findings reveal that expert and individual assessments align more closely with high-quality courtyards, suggesting that shared perceptions can help guide enhancements in courtyard design. This shared perception can also be influenced by environmental conditions. For instance, the quality of office environments is assessed through a comprehensive Indoor Environmental Quality (IEQ) tool, which evaluates factors such as indoor air quality, thermal comfort, visual comfort, acoustic quality, and electromagnetic pollution [184]. This study provides a quantitative, comprehensive approach to assessing IEQ, allowing for objective comparisons across different office environments and emphasizing the role of environmental factors in shaping user experience. Perception can also be shaped from a social perspective. Research on commercial public spaces has shown that social perception has a strong positive effect on both regional and personal identity, while spatial perception produces mixed effects [185].

This study focuses on the Physical Landscape, Social, and Emotional and Spiritual dimensions of placeness, emphasizing how social factors contribute to a sense of identity within urban spaces.

Historic centers also come under public facilities and a study conducted in [186] emphasizes the importance of balancing preservation efforts with community preferences. The study focused on three quality factors: historical value, aesthetic value, and functional value and identified varying levels of integrity across different zones. In short, different public spaces like campuses, playgrounds, courtyards, offices, and commercial areas, each have unique qualities that affect how people experience and benefit from them. These studies show that thoughtful design, accessibility, and a focus on social and environmental factors can make a real difference, helping these spaces better serve their communities.

### *4.7. Cross-Cutting Quality Factors*

Some factors influence the quality of all types of public spaces, regardless of their specific form or function. These factors are not limited to a single category but apply across different spatial types. Cross-cutting factors include accessibility, inclusiveness, safety, publicness, comfort, sustainability, and maintenance. These aspects shape how people interact with public spaces and determine their usability and effectiveness. For example, accessibility is a key factor in urban streets, parks, and public facilities, ensuring that spaces are open to all users. Safety plays a role in streets, waterfronts, and open spaces, affecting how people perceive and use these areas. By analyzing these cross-cutting factors, it is possible to identify common challenges and considerations in public space design and management. Understanding these shared factors helps in creating a consistent and adaptable approach for assessing and improving public spaces across different urban contexts.

## 5. Analysis of Quality Factor Patterns Across Public Space Types

This analysis synthesizes quality factors identified across 159 research studies examining five distinct public space typologies. The systematic review reveals both universal design principles and space-specific characteristics that define public space effectiveness.

Figure 2 presents the distribution of studies across space types, establishing the empirical foundation for this analysis. The dataset comprises 62 studies on urban spaces, 40 on open spaces, 35 on parks and waterfronts, 12 on public facilities, and 10 on green spaces, providing comprehensive coverage of public space research.

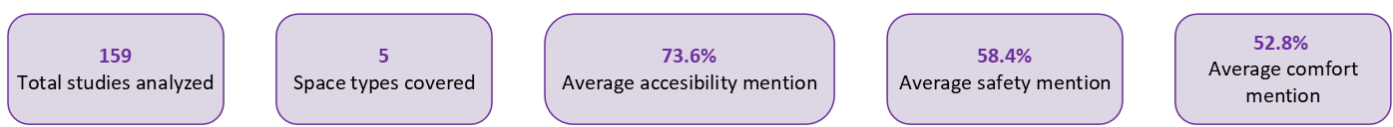

**Figure 2**. Synthesis of 159 research studies across five public space types.

The identification of universal quality factors across all space types, is demonstrated in Figure 3. Accessibility emerges as the predominant factor (73.6% average mention), followed by safety/security (58.4%) and comfort (52.8%). These findings align with established universal design principles and indicate a convergent understanding of fundamental public space requirements across diverse contexts.

Space-specific quality factor profiles are illustrated in Figure 4, revealing distinct prioritization patterns. While accessibility maintains primacy across all typologies, significant variations emerge: open spaces demonstrate highest comfort emphasis (70%), parks and waterfronts prioritize activities (60%), green spaces emphasize aesthetics and natural elements (70% and 60% respectively), and public facilities uniquely prioritize indoor environment quality (41.7%). These variations reflect the functional specialization and user requirements inherent to each space type.

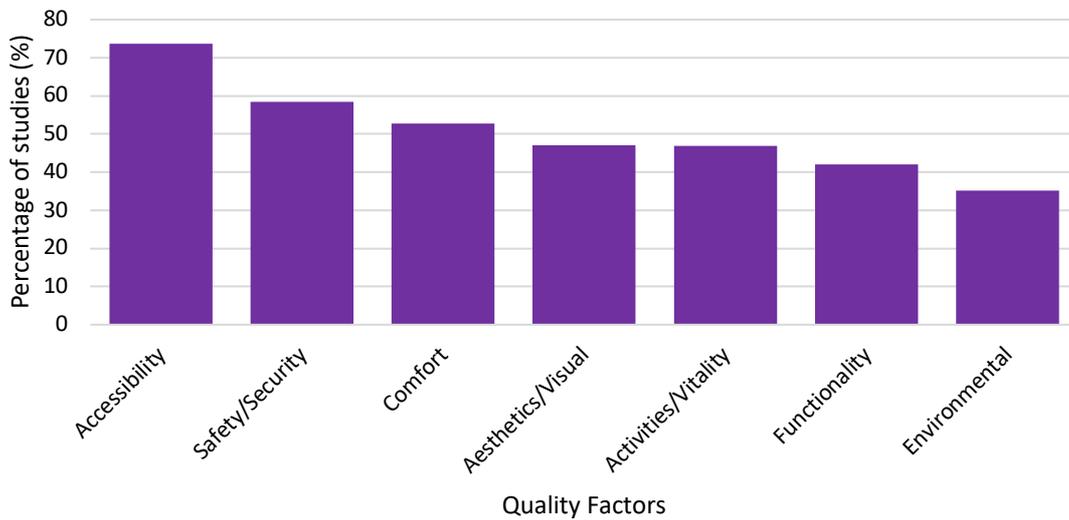

**Figure 3**. Universal Quality Factors Across All Space Types

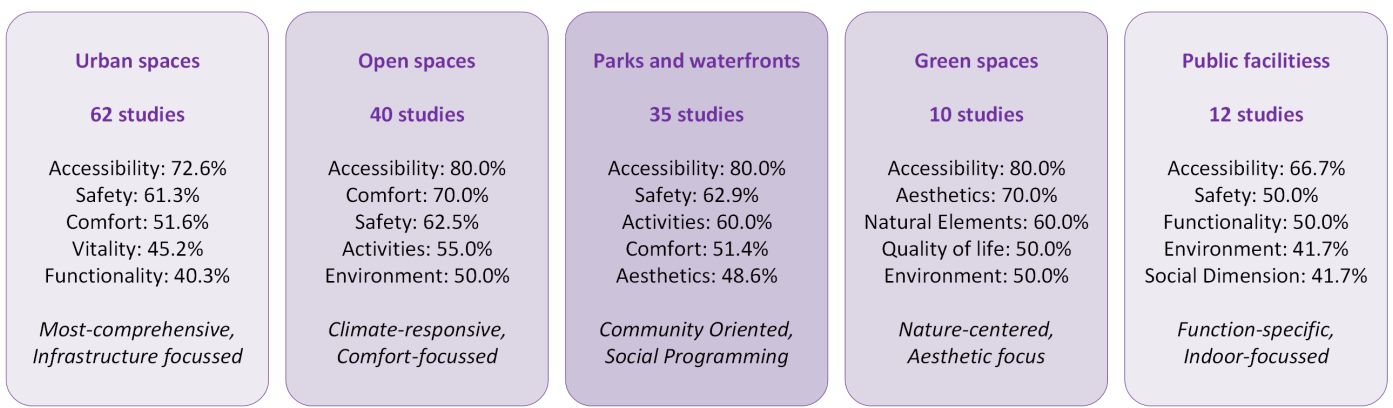

**Figure 4**. Quality factor profiles by space type.

Figure 5 provides a multidimensional comparison of quality factors across space types through radar visualization. The analysis reveals distinct quality factor "profiles" for each space type, with urban spaces showing balanced emphasis across dimensions, green spaces exhibiting peaks in aesthetic factors, and parks/waterfronts demonstrating superiority in social programming metrics.

Figure 6 presents the distribution of quality factor categories across all space types, revealing the relative emphasis placed on different dimensional aspects. User experience factors account for the largest proportion (38% average), followed by physical infrastructure (31%), social interaction (26%), environmental considerations (18%), management aspects (12%), and cultural factors (8%). This categorical analysis demonstrates the predominance of human-centered design considerations while highlighting the supporting role of infrastructure and environmental factors. As illustrated in Figure 6, each public space type demonstrates a distinct 'quality fingerprint' when visualized through radar analysis. While accessibility maintains a consistent high priority across all space types, the remaining quality factors show notable variation. Open spaces exhibit greater emphasis on comfort, parks and waterfronts prioritize activities, green spaces show pronounced peaks in aesthetics and natural elements, and public facilities uniquely prioritize safety and indoor environment quality. This visualization effectively captures the typology-specific prioritization patterns that emerge from our analysis of 159 studies, selected through a rigorous PRISMA-guided process (Figure 1).

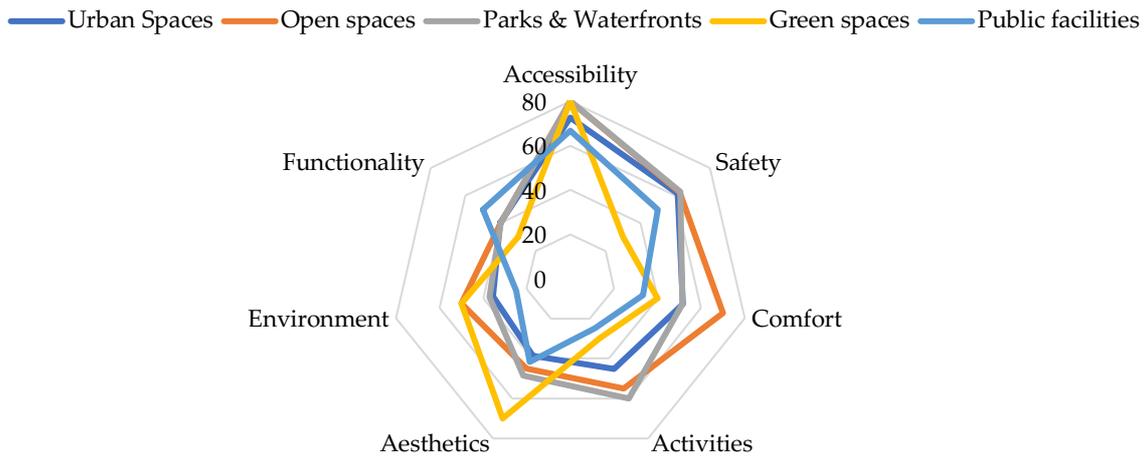

**Figure 5.** Multi-space type quality factor comparison.

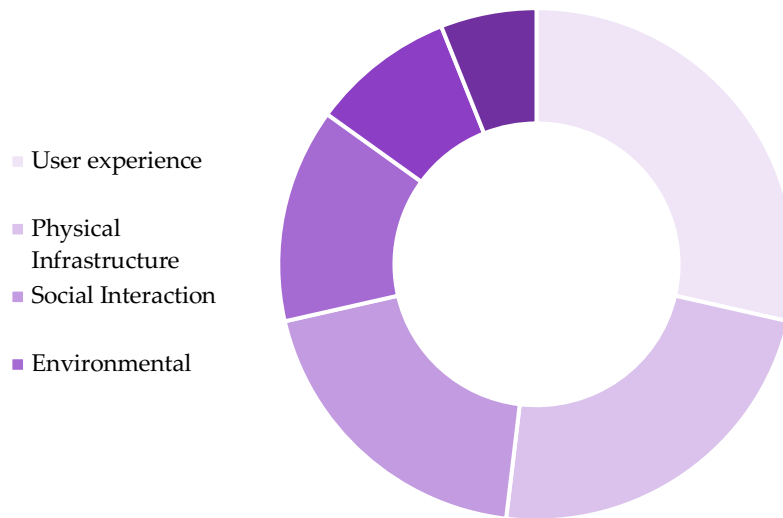

**Figure 6.** Quality factor categories across all space types.

The synthesis of key patterns is presented in Figure 7, which categorizes findings into four analytical domains: universal priorities applicable across all space types, space-specific emphases that differentiate each typology, emerging trends in contemporary practice, and management implications for operational sustainability. The analysis confirms that while accessibility, safety, and comfort constitute universal requirements, specialized factors define space type identity and function.

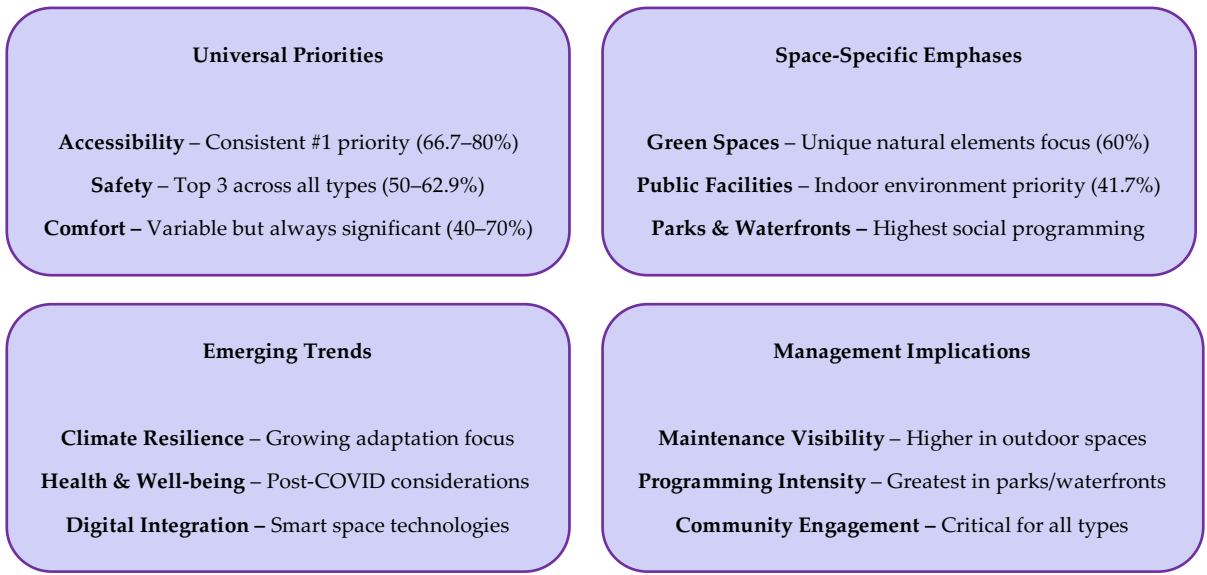

**Figure 7.** Key insights and patterns.

Table 1 documents critical factor interdependencies, revealing a hierarchical structure in quality factor relationships. Accessibility functions as a foundational enabler for all other factors, safety serves as a prerequisite for space utilization, and comfort determines engagement duration and quality. These interdependencies suggest that effective public space design requires sequential consideration of factors based on their enabling relationships.

**Table 1.** Critical factor interdependencies.

| Primary Factor | Enables/Required For | Relationship Type | Impact Level |
|---|---|---|---|
| Accessibility | All other factors | Foundational | Critical |
| Safety | Usage and activities | Prerequisite | High |
| Comfort | Duration of stay | Enabling | High |
| Maintenance | Quality perception | Supporting | Medium |
| Activities | Social interaction | Synergistic | Medium |

The analysis demonstrates that public space quality assessment must account for both universal design principles and typology-specific requirements. While certain factors achieve cross-typological importance, optimal public space development necessitates understanding the unique functional demands and user characteristics of each space type.

## 6. Discussion

The analysis of quality factors across different public space types reveals important patterns that extend beyond mere identification of assessment criteria. The findings highlight both universal human needs and context-specific requirements that should inform more comprehensive approaches to public space evaluation.

The emergence of accessibility, safety/security, and comfort as predominant factors across all space types reflects fundamental human needs in public environments. This convergence suggests a foundation for universal design standards that transcend specific typologies. However, the distinct quality profiles for different space types with open spaces emphasizing comfort, parks prioritizing activities, green spaces focusing on aesthetics and natural elements, and public facilities highlighting indoor quality demonstrate that effective assessment must also address contextual specialization. This duality challenges the notion of one-size-fits-all evaluation frameworks and suggests that public space assessment requires both universal standards and typology-specific criteria. The hierarchical relationship between factors, where accessibility enables other qualities and safety serves as a prerequisite for utilization, provides important guidance for prioritizing interventions and resource allocation.

Current approaches to public space assessment suffer from significant limitations. AI-based studies [187], [188], [189], [190], [191], [192], [193], [194], [195], [196], [197] often prioritize computationally convenient factors over comprehensive evaluation, while domain-specific research remains confined within disciplinary silos. Many assessments rely excessively on user perceptions without integrating objective measurements, resulting in partial understanding of complex public environments. The fragmentation in assessment methodologies underscores the need for integrated frameworks that can address public spaces as complete urban systems. Such frameworks should establish core universal criteria while providing space-specific modules, include temporal dimensions, integrate objective and subjective measures, consider cultural variations, and transcend disciplinary boundaries. This research contributes to the field by providing the first comprehensive synthesis of quality factors across diverse public space typologies. By identifying both universal principles and space-specific requirements, it establishes a foundation for more nuanced assessment approaches. The hierarchical framework of factor relationships offers insights into how different dimensions interact and influence overall space performance. For practice, the findings suggest that urban planners and designers should prioritize universal accessibility, safety, and comfort while tailoring other interventions to specific space types. Policymakers should develop assessment frameworks that balance standardization with contextual flexibility.

Future research should focus on developing and validating comprehensive assessment frameworks, investigating the temporal dynamics of public space quality, exploring the impact of digital technologies, and examining relationships with broader urban sustainability goals.

## 7. Conclusion

This study has systematically analysed public space quality assessment approaches across five spatial typologies through examination of 159 research studies. Our findings reveal both universal quality factors that transcend spatial types and specialized factors that define unique requirements of different public environments. The identification of accessibility, safety/security, and comfort as foundational requirements across all public space types establishes a universal baseline for quality assessment. Meanwhile, the distinct quality priorities for different space types highlights the importance of contextual specialization. The hierarchical structure of quality factors, where accessibility enables other qualities, safety serves as a prerequisite for utilization, and comfort determines engagement quality, provides important insights for design prioritization. Our analysis identifies critical limitations in current assessment approaches, including AI studies focused on easily quantifiable factors, domain-specific research confined within disciplinary boundaries, and overreliance on subjective perceptions without objective measures. These limitations underscore the need for integrated, multidimensional data-driven analysis and framework development that address public spaces as complex urban systems.

The implications for practice include prioritizing universal accessibility, safety, and comfort while tailoring other interventions to specific space types. Policymakers should develop balanced assessment frameworks that ensure both comparability across spaces and sensitivity to local conditions. Researchers should work toward integrated methodologies that combine objective measures with subjective perceptions. Public spaces play a vital role in urban life, contributing to social cohesion, environmental quality, and community wellbeing. This research provides a foundation for more integrated approaches to public space assessment, moving beyond fragmented methodologies toward comprehensive frameworks that acknowledge the complexity of public urban environments. By synthesizing insights from multiple perspectives and space types, this study contributes to the development of more effective assessment tools that can support the creation of high-quality public spaces for all.

**Author Contributions: S.**T., M.J., and S.D., conceived and designed the work. S.D., and R.A. critically revised the article. All authors have read and agreed to the published version of this manuscript.

**Funding:** This research was funded by the United Arab Emirates University through a UAEU Strategic Research Grant G00003676 (Fund No: 12R136), via the Big Data Analytics Center.

**Acknowledgments:** The authors thank the United Arab Emirates University for supporting this work through the UAEU Strategic Research Grant (G00003676) and the Abu Dhabi International Virtual Research Institute for Food Security in the Drylands (G00004017 - VRI-FS 120-21)

**Conflicts of Interest:** The authors declare no conflicts of interest. The funders had no role in the design of the study; in the collection, analyses, or interpretation of data; in the writing of the manuscript; or in the decision to publish the results.

## Abbreviations

| | |
|---|---|
| BIM | Building Information Modeling |
| DQIs | Design Quality Indicators |
| GPSI | Good Public Space Index |
| IDU | Intensity and Diversity of Use |
| IEQ | Indoor Environmental Quality |
| LRT | Light Rail Transit |

MCDA     Multi-Criteria Decision Analysis

POSI     Public Open Space Index

PSI     Public Space Index

PSQI     Public Space Quality Index

WAT     Walkability Assessment Tool